%% file: article.tex
\documentclass[fedsstyle, oneside, onecolumn, Numbered]{jnl}

\usepackage{graphicx}
\usepackage{amsmath,amssymb,amsfonts}
\usepackage[title]{appendix}
\usepackage{manyfoot}
\usepackage{booktabs}
\usepackage{bibunits}
\usepackage{titletoc}
\usepackage{lmodern}
\usepackage{transparent}

\defaultbibliographystyle{fedsstyle}
\hypersetup{urlcolor=black}

\title{Deep, label-free, volumetric lifetime microscopy via low-noise, single pulse sampling}

\author[1,2]{\fnm{Jack} \sur{Tomkiewicz}}\equalcont{}
\author[1,2]{\fnm{Federico} \sur{Presutti}} \equalcont{These authors contributed equally to this work.
\hfill\makebox[0pt][l]{\scalebox{0.0}{\texttransparent{0}{\nolinkurl{fpresutt@mit.edu}}}}}
\author[1,2]{\fnm{Tade} \sur{Marozsak}}
\author[3,4]{\fnm{Jia} \sur{Zhao}}
\author[3,4]{\fnm{Hatoon} \sur{Baazim}}
\author[3,4]{\fnm{Hernandez} \sur{Moura-Silva}}
\author[3]{\fnm{Eliezer} \sur{Calo}}
\author[1,2]{\fnm{Sixian} \sur{You}}

\affil[1]{%
 \orgdiv{Research Laboratory of Electronics},
 \orgname{Massachusetts Institute of Technology},
 \orgaddress{%
   \state{MA}%
   , \country{USA}}%
}
\affil[2]{%
 \orgdiv{Department of Electrical Engineering and Computer Science},
 \orgname{Massachusetts Institute of Technology},
 \orgaddress{%
   \state{MA}%
   , \country{USA}}%
}
\affil[3]{%
 \orgdiv{Department of Biology},
 \orgname{Massachusetts Institute of Technology},
 \orgaddress{%
   \state{MA}%
   , \country{USA}}%
}
\affil[4]{%
 \orgdiv{Ragon Institute}
 \orgname{of MGH, MIT and Harvard},
 \orgaddress{%
   \state{MA}%
   , \country{USA}}%
}

\abstract{
Fluorescence lifetime imaging (FLIM) provides label-free metabolic contrast, but conventional time-correlated single-photon counting requires thousands of excitation pulses per pixel to avoid pile-up distortion, limiting imaging speed and increasing motion artifacts and photodamage.
Here, we develop single-pulse analog sampling that combines sub-photoelectron read noise with a jitter-immune, frequency-domain estimator for accurate lifetime measurement at low photon counts.
Single low-flux traces achieve a photon economy of $F < 1.5$ at 100--150 detected photons, enabling accurate lifetime estimation from a single excitation pulse in label-free, volumetric samples.
In in vivo zebrafish eyes, single-shot acquisition recovered the lifetime content of 20-pulse averaging at one-twentieth the exposure, resolving endogenous lipofuscin heterogeneity.
In intact mouse kidney, scattering and reduced photon throughput impaired lifetime contrast; this was mitigated by lightweight denoisers that enabled label-free lifetime imaging to 200~{\textmu}m depth at one pulse per pixel.
These results establish single-shot FLIM as a route to deeper, faster, motion-tolerant, low-photodamage lifetime imaging.
}

\begin{document}

\maketitle


\input{text_introduction}

\section*{Results}
\input{text_validation}
\input{text_zebrafish}
\input{text_kidney}

\input{text_discussion}

\input{text_methods}

\bmhead{Funding}
This work was supported by National Science Foundation CAREER Award (2339338), National Institutes of Health (6956156 - 1RD01DK147407-01), CZI Dynamic Imaging from the Chan Zuckerberg donor-advised fund through the Silicon Valley Community Foundation, the Manton Foundation, the Fairbairn Menstruation Science Fund, Novo Nordisk Research Development, and MIT HEALS.

\bmhead{Acknowledgments}
We thank Timothy Weber for sharing his software interface for the digitizer.
We thank Tathagata Banerjee and Li-Yu Yu for helpful comments regarding the manuscript. We thank Mingze Jiang for her help investigating various denoising methodologies.



\vspace{3\baselineskip}

\bmhead{Author Contributions}
\phantom{.} \\
\textbf{J.T.}: Conceptualization, Data curation, Formal analysis, Investigation, Methodology, Project administration, Software, Supervision, Validation, Visualization, Writing. \\
\textbf{F.P.}: Conceptualization, Data curation, Formal analysis, Investigation, Methodology, Project administration, Software, Supervision, Validation, Visualization, Writing. \\
\textbf{T.M.}: Investigation, Data curation. \\
\textbf{J.Z.}: Methodology and Resources. \\
\textbf{H.B.}: Methodology and Resources. \\
\textbf{H.M-S.}: Investigation and Resources. \\
\textbf{E.C.}: Conceptualization, Investigation, Methodology, Resources, Validation. \\
\textbf{S.Y.}: Conceptualization, Funding acquisition, Methodology, Project administration, Supervision, Visualization, Writing.

\bmhead{Code availability}
An implementation of the fitting program can be found at \href{https://github.com/You-Lab-MIT/fdexpfit}{github.com/You-Lab-MIT/fdexpfit}.

\input{text_supplementary}



\bibliography{refs}

\end{document}

%% file: text_introduction.tex
\phantomsection\addcontentsline{toc}{section}{Introduction}

Fluorescence lifetime imaging microscopy (FLIM) identifies fluorescent species and reveals molecular states that fluorescence intensity alone cannot distinguish.
Because fluorescence lifetime is influenced by the local molecular environment while remaining comparatively insensitive to excitation power and light collection efficiency, it can indirectly measure temperature, viscosity, ion concentration, and protein--fluorophore interactions \cite{okabe_intracellular_2012, hosny_mapping_2013, wallrabe_imaging_2005}.
This distinction is particularly valuable for endogenous fluorophores.
Free and protein-bound NAD(P)H, for example, exhibit distinct fluorescence decay profiles \cite{lakowicz_fluorescence_1992, song_visualizing_2024}, enabling autofluorescence lifetime imaging to report cellular metabolic state without exogenous labels.
Such measurements have classified T-cell activation \cite{walsh_classification_2021}, monitored metabolic fitness during CAR-T-cell manufacturing \cite{pham_label-free_2026}, delineated tumor margins \cite{alfonso-garcia_vivo_2023}, and predicted the remaining lifespan of individual \textit{C. Elegans} \cite{morrow_endogenous_2024}.
The advantages of this label-free contrast are especially compelling in living tissues, particularly ones that are deep and scatter light.
In such tissues, labels can perturb the system and three-dimensional biochemical measurements can be difficult to obtain.
Yet this is also the regime in which lifetime imaging is most challenging.

\begin{figure}[!htbp]
\centering
\includegraphics[width=\normaltextwidth]{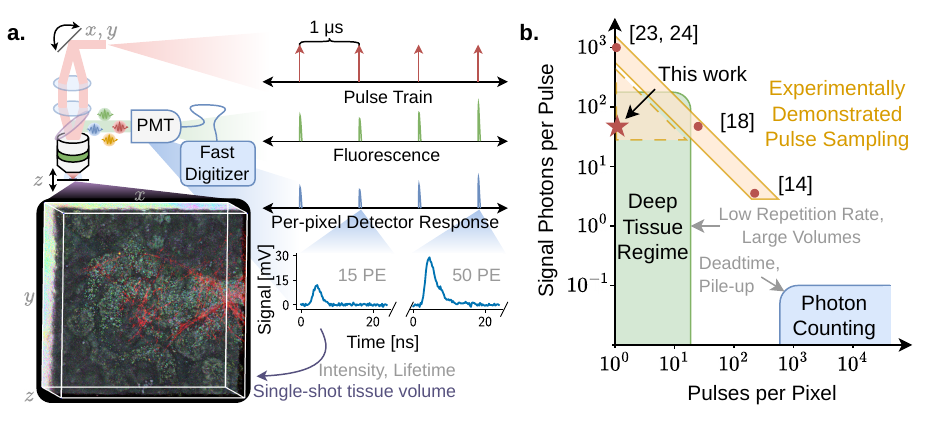}
\caption{
\textbf{Single-shot fluorescence lifetime imaging in tissue.}
\textbf{a.} Schematic of the single-shot FLIM approach.
A scanned multiphoton excitation beam generates a fluorescence signal that is detected by a PMT and temporally sampled by a high-speed digitizer.
The resulting PMT waveform simultaneously encodes fluorescence intensity and lifetime, enabling lifetime estimation from as few as a single excitation pulse.
Shown is a 3D rendering of a stack of images acquired from a mouse kidney at increasing depth, where each pixel was obtained from a single laser pulse.
\textbf{b.} Photon economy of fluorescence lifetime imaging methods, showing the number of detected photons per excitation pulse versus the number of excitation pulses required per pixel.
Conventional photon-counting FLIM operates in the low-photon-rate, high-repetition regime, whereas experimentally demonstrated pulse-sampling approaches utilize fewer pulses, but with high photon-rates.
This work operates in the single-pulse, deep-tissue regime, enabling FLIM from a single excitation pulse per pixel.
}
\label{fig:intro}
\end{figure}

The standardization and commercialization of time-correlated single-photon counting (TCSPC) were pivotal in making FLIM a sensitive, quantitative, and broadly accessible measurement.
Turnkey lifetime detection is now a standard option on mainstream confocal platforms, an established readout in high-content FRET screening \cite{kumar_flim_2011}, and the basis of fluorescence lifetime imaging ophthalmoscopy, which has carried label-free retinal lifetime measurement into multi-site clinical research on age-related macular degeneration and diabetic retinopathy \cite{sauer_fluorescence_2019}.
TCSPC achieves near-ideal precision for a given photon budget by building an arrival-time histogram one photon at a time, but conventional implementations restrict detection to a small fraction of excitation cycles to avoid pile-up; consequently, even an estimate from 100--150 photons typically requires thousands of pulses \cite{becker_fluorescence_2006, hirvonen_fast_2020}.
Multi-event electronics, parallel detectors, and computational photon counting raise this limit \cite{sorrells_computational_2022}, but the measurement remains assembled across repeated excitation cycles.
This becomes especially restrictive in deep-tissue imaging, where the laser repetition rate is often reduced from 80 MHz to a few megahertz or even sub-megahertz to allow higher pulse energies at a given average power \cite{hontani_multicolor_2021, redlich_high-pulse-energy_2021, boguslawski_increasing_2024}. 
For these applications, the low excitation rate, combined with the limited detection rate of conventional TCSPC due to pile-up constraints, can substantially increase acquisition times. 
Reliably recovering the lifetime from only a few or even a single excitation pulse – similarly to measuring fluorescence intensity – is therefore highly attractive.

Advances in acquisition methods in the form of analog pulse sampling \cite{moon_analog_2009, giacomelli_rapid_2015, bower_high-speed_2018} provide a route to capture the fluorescent decay profile following individual excitation pulses by reading the photodetector's electronic output with a high-speed digitizer, thus avoiding the photon rate limitations necessary to avoid pileup.
This enables the readout of hundreds of photons per shot (Fig.~\ref{fig:intro}a).
Since the photons used to form a fluorescence intensity image also carry information about their time of arrival, lifetime information can, in principle, be extracted from the same photons. 
Thus, lifetime imaging need not impose a fundamental penalty in imaging speed or photon budget, although achieving a given lifetime precision may require additional photons in practice.
Pulse sampling has already been demonstrated at 7.3 megapixels per second, matching resonant-scanned intensity multiphoton imaging \cite{giacomelli_rapid_2015}.
Despite this, it has yet to be adopted for deep and label-free imaging where the number of measured photons per pulse is typically quite low.
The gap to adoption is in the lifetime measurement itself.
Unlike in photon counting methods, the waveform is not thresholded and hence the usual drawbacks of analog methods apply: read noise, baseline drift, timing jitter, and pulse-to-pulse variations in the detector response all can directly affect the lifetime estimate.
A recent article reported a pulse sampling photon economy (the figure of merit for the efficiency of lifetime estimation, discussed below \cite{GerritsenAsselbergs2002, BallewDemas1989}) of $F \approx 2.4\text{--}3$ (accounting for detector quantum efficiency) -- compared with $F \approx 1\text{--}1.3$ for TCSPC (the shot-noise limited optimal photon economy is $F=1$), and attributed much of the difference to laser-pulse variability \cite{zhou_pulse-sampling_2023}.
Another work reported $F=1.2$ for a long lifetime fluorophore ($\tau = 3.24$\,ns), but $F=2.5$ for short lifetime species ($\tau = 0.9$\,ns) \cite{moon_analog_2009}.
Pulse sampling has therefore been applied predominantly to bright, labeled emitters \cite{eibl_single_2017}, pulse-averaged acquisition \cite{ryu_real-time_2018}, or two-bin lifetime prediction \cite{li_single-shot_2018}.
Crucially, this inefficiency compounds in deep-tissue, label-free imaging, where endogenous fluorescence is weak and low-repetition-rate excitation concentrates the available average power into fewer and more energetic pulses \cite{liu_deep_2024, lecoq_wide_2019}.
This combination of acquisition volume and the number of pulses required for lifetime estimation presents a significant technical hurdle for deep label-free lifetime imaging (Fig.~\ref{fig:intro}b).
However, there is no fundamental reason why pulse sampling cannot match the performance of counting-based techniques: reaching the robustness and immunity to noise of thresholding techniques is ultimately a matter of hardware fidelity and optimal data analysis.
 
In this work, we demonstrate single-shot label-free fluorescence lifetime imaging and bring it near its theoretical maximum photon efficiency by combining a front end whose read noise falls below one photoelectron with a jitter-immune, baseline-corrected, frequency-domain estimator.
With this method, single low-flux traces reach $F < 1.5$ at only 100--150 photons, measured on individual raw traces with no pulse averaging, no spatial context, and no learned spatial prior.
Since a single multiphoton excitation pulse can deliver up to hundreds of photons, it is possible to obtain a relatively accurate lifetime estimate in a single shot.
In contrast, photon counting methods typically use hundreds to thousands of pulses to measure the same number of photons and thus reach the same accuracy.
For signals that fall below the range in which a single excitation pulse supports accurate lifetime estimation, we further introduce lightweight denoisers that reduce low-photon lifetime bias either independently at each pixel or by incorporating detector-readout information from a local spatial neighborhood.
To our knowledge, these results constitute the first demonstration of single-pulse-per-pixel, label-free multiphoton FLIM at endogenous-fluorescence photon levels.

We demonstrate this capability in two complementary biological regimes.
In a living zebrafish (\textit{Danio rerio}) eye, the fluorescent output yields high SNR PMT output traces which enable single-shot acquisition.
We found the single-shot lifetimes to closely match the lifetimes of a highly averaged acquisition, all while removing the potential motion sensitivity and cumulative light dose that averaging imposes on a living animal.
Furthermore, this single-shot acquisition is able to resolve lifetime heterogeneity within the retinal pigment epithelium that intensity contrast alone does not reveal.
In an intact and fresh mouse kidney, weaker fluorescence and the densely scattering cortex limit photon collection.
While lifetime contrast between various regions is still maintained, there are some layers with insufficiently low photon counts that the lifetime estimates become significantly corrupted by noise.
To fix this, we evaluate three denoising methods of varying complexity: Gaussian filtering, bilateral filtering, and Bayesian denoising.

These experiments establish the photon-budget boundary of single-shot lifetime imaging across transparent and strongly scattering tissues and show how far the technology can be taken.
Requiring only the addition of a 4 GS/s digitizer to an otherwise conventional point-scanning microscope, this approach makes near-counting photon efficiency available down to one pulse per pixel, with potential value for volumetric and longitudinal metabolic imaging of organoids and engineered tissue \cite{qian_label-free_2021}, intraoperative label-free margin assessment \cite{alfonso-garcia_vivo_2023}, and retinal imaging \cite{sauer_fluorescence_2019}.

%% file: text_validation.tex
\subsection*{Characterization of lifetime measurement via pulse sampling}\label{sec:validation}
\addcontentsline{toc}{section}{\nameref{sec:validation}}

\begin{figure}[!htbp]
\centering
\includegraphics[width=\normaltextwidth]{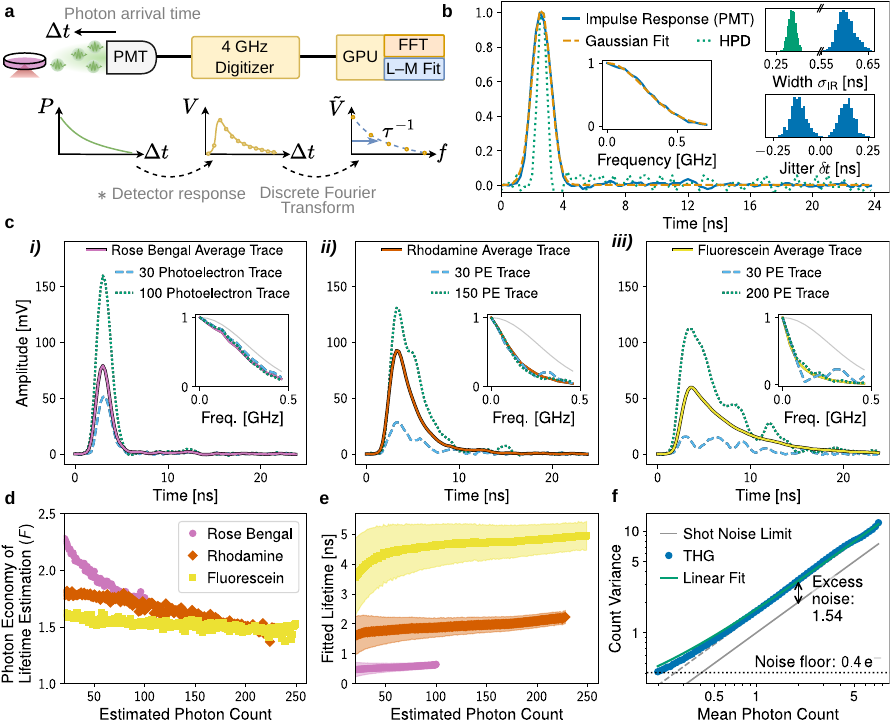}
\caption{
\textbf{Characterization of multiphoton fluorescence lifetime measurements at low photon counts, via pulse sampling and curve fitting.}
\textbf{a.} Fluorescence lifetime temporal trace acquisition.
The microscope acquisition hardware consists of a photomultiplier tube (PMT), whose output undergoes analog to digital conversion at 4\,GHz.
Temporal traces of the fluorescence decay profile are acquired for each pixel, and analyzed to obtain an intensity and an estimate of the lifetime ($\tau$).
The acquired signals are a combination of the arrival-time probability ($P$) of the emitted photons collected from the sample, convolved with the instrument response, and the digitization of these analogue electronic signals.
Finally, a Levenberg--Marquardt (L--M) type fitting is performed, computed in parallel on the GPU, in the frequency domain, following a fast Fourier transform (FFT).
\textbf{b.} Instrument impulse response profile.
A single trace is shown 
of a zero-lifetime measurement.
The response from a hybrid photodetector (HPD) is shown for comparison.
A Gaussian fit to the trace is shown, which was used for curve fitting.
Insets: Fourier transform of the trace and the fitted Gaussian; distribution histogram of the Gaussian $\sigma_\text{IR}$ (including for HPD), and deviation of the center $\delta t$, over 1000 traces.
\textbf{c.} Example traces of three calibration dyes: i) Rose Bengal, ii) rhodamine and iii) fluorescein.
For each sample, averaged traces and two individual traces -- at low and high signal intensities -- are shown.
The insets show the same traces in the frequency domain, on which the fits were performed.
The gray line represents the impulse response spectrum.
\textbf{d.} Photon economy of the lifetime estimation ($F$) for frequency fitting on the three dyes: how precisely and efficiently the lifetime is inferred from repeated acquisitions on the same sample, with respect to the number of photons detected.
A value of 1 saturates this metric.
The traces were binned based on the estimated photon count (see \hyperref[methods:penumber]{Methods}).
\textbf{e.} Fluorescence lifetime mean and standard deviation of the three dyes, with respect to number of detected photons.
\textbf{f.} Photon transfer curve.
The sub-photoelectron noise floor characterizes the Gaussian noise and the deviation from the shot noise limit characterizes the detector excess noise.
}
\label{fig:validation}
\end{figure}

Our implementation digitizes the PMT output at 4\,GS/s and fits each transient in the frequency domain with a real-time (0.35 \textmu s/trace) GPU curve-fitting implementation (Fig.~\ref{fig:validation}a and \hyperref[methods:fitting]{Methods}).
We characterized the instrument response from third-harmonic generation at a water--glass interface (a zero-lifetime reference), obtaining a Gaussian with a standard deviation of $\sigma_\text{IR}$ = 0.6\,ns that is reproducible to $\Delta\sigma_\text{IR}$ = 20 ps across 1,000 single-shot measurements (Fig.~\ref{fig:validation}b); this width sets the 0.5\,GHz fitting bandwidth, which can further be seen by looking at the fourier domain spectra in Fig.~\ref{fig:validation}c.
We found that the time jitter distribution was bimodal, with separation equal to the digitizer clock period and comparable widths.
To evaluate the detector noise, we measured the photon-transfer curve (Fig.~\ref{fig:validation}f), which quantifies the signal intensity mean--variance relationship.
The measured variance was well described as proportional to the mean plus a constant read-noise floor.
The slope suggests an excess noise factor of 1.5 above shot noise, likely dominated by the detector's gain variation.
The noise floor was equivalent to approximately 0.4 photoelectrons; in other words, the electronic read noise is less than the signal generated by a single detected photon.
Consequently, once a temporal sample contains on the order of one photoelectron or more, the measurement is dominated by photon (shot) noise rather than electronic noise.
This demonstrates that the acquisition hardware already operates close to the fundamental photon-noise limit, providing the conditions necessary for achieving near-shot-noise photon economy in the lifetime estimation, as defined below.

Even with a near-shot-noise hardware apparatus, recovering a lifetime from a single-pulse low-flux trace remains challenging for several reasons.
First, temporal jitter between the laser and digitizer causes a slight shift in arrival time for each pulse, which is exacerbated by the finite sampling rate of the digitizer.
Second, due to the digitizer electronics, each trace has a slightly different DC baseline, complicating the analysis.
Third, at low signal intensities the few detected photons are less likely to include late arrivals, and these later arrivals will be sparse or more likely to be buried under the electronic noise floor.
This distorts the trace away from a monotonic decay profile, and biases the lifetime estimation downwards.
This penalty deepens as the lifetime increases.
These first two effects -- jitter and baseline drift -- however, tend to strongly penalize shorter lifetimes, where the signal is highly localized in time and where a small timing error in absolute terms can result in a large relative lifetime error.
Lastly, when the lifetime $\tau$ is comparable to or below the instrument response (IR) time $\sigma_\text{IR}$, the measurement is bottlenecked by the shot-to-shot spread of the IR itself.
Specifically, as the timescales add in quadrature, the overall response time $T$ for short lifetimes is given as
\begin{equation}
    T = \sqrt{\sigma_\text{IR}^2 + \tau^2},
\end{equation}
by modeling the response as an exponentially modified Gaussian.
When $\tau < \sigma_\text{IR}$, $T$ broadens weakly with $\tau$, meaning that the width variation $\Delta\sigma_\text{IR}$ may have a greater effect on the response width than the lifetime itself.

These complications can be mitigated by trace averaging and brighter signals.
Together these make single-shot traces differ markedly in amplitude, offset, and timing (Fig.~\ref{fig:validation}c), which is likely what confined earlier analog-signal methods to bright emitters or multi-pulse averaging.

We found that adopting Fourier-domain amplitude fitting for single-shot sampling can mitigate some of these issues, and particularly benefits the measurement of short lifetimes.
The insets in Fig.~\ref{fig:validation}c show this directly: traces that look different in time have normalized moduli that overlap across the low-frequency, lifetime-encoding band.
A significant source of variance between individual traces is in the higher frequency content, where the signal vanishes and the noise therefore dominates.
Because the finite photodetector rise time, we restrict the fit to a low-frequency band ($<$0.5\,GHz) which still contains the temporal information of the response. 
This discards the high-frequency read noise while keeping the full lifetime information, which is carried in how fast the amplitude rolls off with frequency.
The model assumes a single-exponential decay: for multiexponential decays, the fitted lifetime represents the average sample lifetimes contained within the pixel, rather than representing a true molecular decay or indicating a biochemical state.
A time shift multiplies the spectrum only by a phase factor $\exp(-i\omega t_0)$, so the Fourier domain amplitude is invariant to timing jitter or window shifts.
With these steps (see \hyperref[methods:fitting]{Methods}), lifetime readout becomes less sensitive to timing jitter and high-frequency noise -- the combination that makes single-shot, low-photon estimation feasible.
The integration of near-shot-noise-limit pulse sampling hardware and a noise-robust estimator is what sets the estimator apart from the mean-delay and phasor readouts of prior analog FLIM methods for short-lifetime emitters (see \hyperref[supp:analysis]{Supplementary} for more information).
These methods are built on absolute timing and phase -- the mean delay is the first time-moment, and the phasor rotates by $\delta\omega t_0$ under a time shift -- and are therefore corrupted by exactly the shot-to-shot jitter that single-shot operation cannot average away \cite{moon_analog_2009, digman_phasor_2008}.

To establish accuracy, we imaged three reference dyes spanning the biological lifetime range: rose bengal, rhodamine B, and fluorescein ($\tau \approx$ 0.5, 1.7, and 5\,ns respectively) (Fig.~\ref{fig:validation}c).
Fitted lifetimes matched reference values across two orders of magnitude in photon number (Fig.~\ref{fig:validation}d).
We measured this precision using the commonly used figure of merit known as photon economy for lifetime estimation.
The photon economy figure of merit is given as
\begin{equation}
F = \frac{\sigma_{\tau}}{\sigma_0}
  = \frac{\sigma_{\tau} \sqrt{N}}{\tau} ,
\end{equation}
for which $F = 1$ corresponds to the bound \cite{GerritsenAsselbergs2002}.
The equation tells us that, when saturating the bound, 100 photons are required to achieve estimates within 10\% error, $10^4$ for 1\%, and so forth \cite{kollner_how_1992}.
Here $\tau$ is the lifetime of an exponential decay, $N$ is the number of detected photons, $\sigma_{\tau}$ is the achieved standard error in lifetime estimation, and $\sigma_0=\tau/\sqrt{N}$ is the ideal standard error of the mean for an exponential distribution.
As with shot noise in intensity imaging, the number of photons constrain the certainty of -- or available ``information'' about -- the lifetime, through their individual arrival times.

For our acquisition method, the precision of estimating short and long lifetimes is limited by different physics.
With Rose Bengal, for which the short lifetime is comparable to the instrument response, precision was bottlenecked by the shot-to-shot spread of the IR itself.
Due to this fundamental bottleneck of pulse sampling, we measured a photon economy ranging between 1.7 and 2.3.
The estimation of longer lifetimes (e.g.\ fluorescein) at very low intensities is bottlenecked by the lack of photons sampled to realize a whole decay curve, especially at the tail where the arrival-time probability is lowest. Conversely, at high photon numbers, we measured decay-curve broadening due to detector nonlinearity, which increased the measured lifetime.
In our case, this caused the estimated lifetime to increase by up to hundreds of picoseconds (see \hyperref[supp:nonlinearity]{Supplementary}). 
This effect is reproducible and could be calibrated in practice.
For both fluorescein and rhodamine B, $F$ fell below 1.5 at 100--150 detected photons and decreased further with the photon number as SNR improved (Fig.~\ref{fig:validation}e; well-implemented TCSPC reaches $F \approx$ 1--1.3), which is at least 40\% better than the reported pulse-sampling economy of short lifetime species \cite{zhou_pulse-sampling_2023, moon_analog_2009}. 

Overall, while time-domain methods may be sufficiently noise-resilient or even outperform fitting for longer lifetimes, the fitting scheme we propose greatly benefits the imaging of short-lifetime species (typical for endogenous fluorophores).
See \hyperref[supp:estimation]{Supplementary} for comparison of different fitting and fit-free methods.
Improvements in photodetector technology and calibration are expected to further increase performance and improve the fidelity of single-shot lifetime estimation.
For example, modern, fast, hybrid photodetectors offer a lower excess noise factor and a smaller transit-time spread, which bring the photon-economy factor closer to the counting ideal \cite{hwang_achieving_2021, sorrells_computational_2022}.
While in this work we focus on conventional PMTs due to their ubiquity in multiphoton imaging systems, in the \hyperref[supp:hpd]{Supplementary} we show that hybrid photodetectors can achieve $F \leq 1.3$ using the same methods and samples we have shown here with PMTs.

Because a single multiphoton pulse in our system can deliver as many as 300 photons, a near-shot-noise-limited lifetime estimate can be obtained at every pixel from a single shot -- the photon budget that counting methods would accumulate only over thousands of pulses.
This dramatic increase in lifetime acquisition speed shifts the bottleneck in FLIM from the photon arrival rate to the laser repetition rate.
This enables a new suite of applications for FLIM, including deep-tissue multiphoton acquisition of three-dimensional lifetime stacks.

%% file: text_zebrafish.tex
\subsection*{Single-shot lifetime measurement of an embryonic zebrafish eye}\label{sec:fisheye}
\addcontentsline{toc}{section}{\nameref{sec:fisheye}}

\begin{figure}[!htbp]
\centering\includegraphics[width=\normaltextwidth]{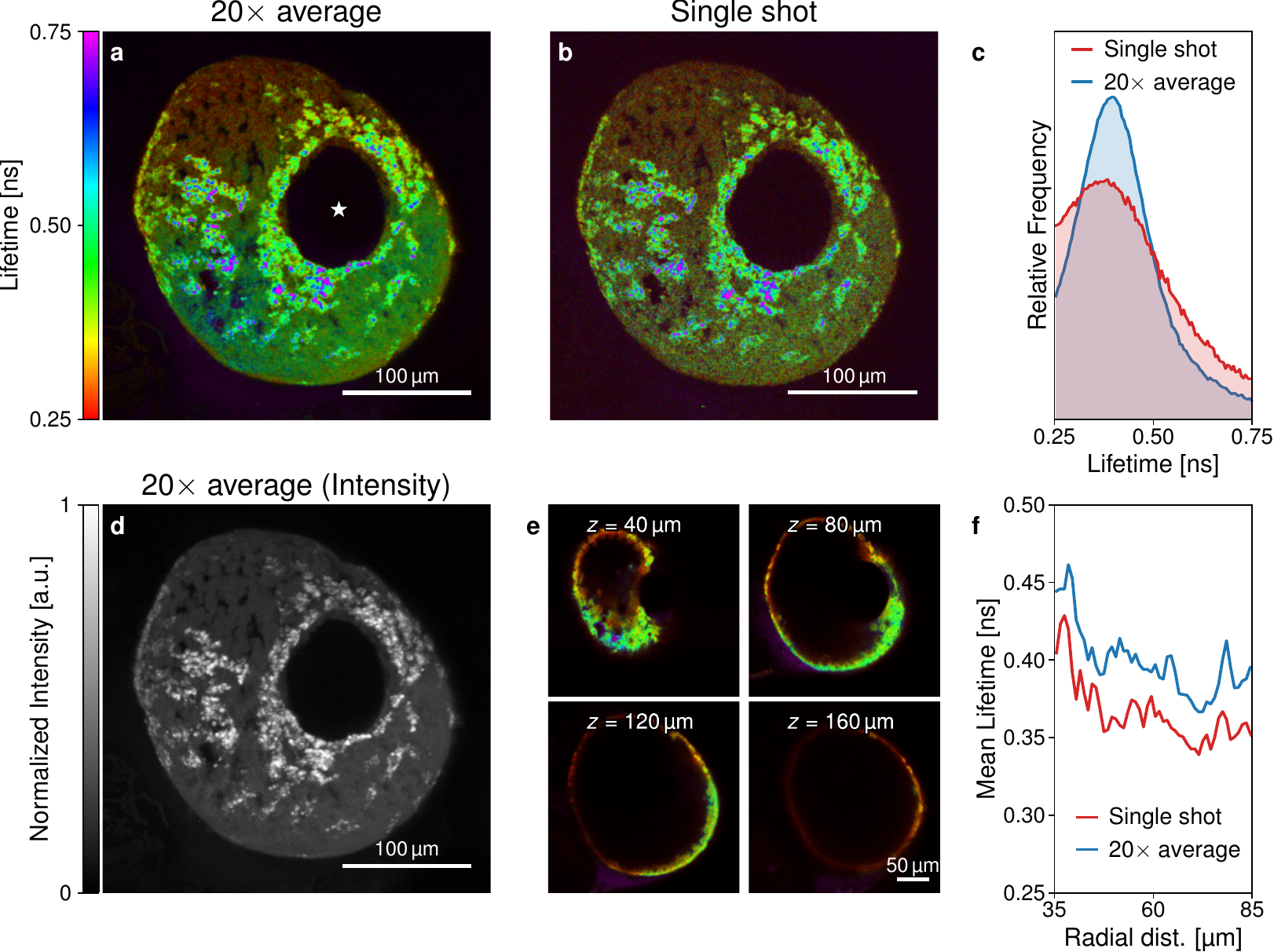}
\caption{
\textbf{Deep FLIM of an embryonic \textit{Danio rerio} (zebrafish) eye.}
\textbf{a.} Maximum intensity projection of a single-shot-image z-stack and \textbf{b} a 20$\times$ averaged z-stack.
\textbf{c.} Intensity-weighted histogram of lifetimes from both datasets.
\textbf{d.} Intensity-only rendering of the 20$\times$ averaged z-stack.
\textbf{e.} Selected 2D slices from the 20$\times$ averaged dataset.
\textbf{f.} Plot of average lifetime versus radial distance from the center of the eye (defined as the location of the star in \textbf{a}).
}
\label{fig:zebrafish}
\end{figure}

We next evaluated whether single-shot acquisition could provide reliable fluorescence-lifetime measurements in the eye of a living zebrafish embryo (\textit{Danio rerio}, Fig.~\ref{fig:zebrafish}).
Because zebrafish retinal architecture and developmental programs are similar to those of humans \cite{howe_zebrafish_2013}, in vivo lifetime imaging of the embryonic eye provides a label-free means to probe biochemical changes during retinal growth and maturation.
However, as the sparse 2D slices in Fig.~\ref{fig:zebrafish}e show, the strongly curved, three-dimensional geometry of the retinal pigment epithelium necessitates volumetric imaging, making acquisition time and excitation exposure critical constraints in vivo.
On our system a twenty-fold-averaged 350 $\times$ 350 $\times$ 200 \textmu m stack takes 21 minutes and delivers twenty times the dose, long enough for photodamage to accumulate.
Based on the photon efficiency demonstrated in Fig.
2, we hypothesized that a single excitation pulse per pixel would provide sufficient signal for lifetime estimation.
If so, a complete three-dimensional lifetime volume could be acquired in approximately 1 minute, the same as the time required for a single-shot volumetric intensity scan while reducing excitation exposure by a factor of 20.

As seen in Fig.~\ref{fig:zebrafish}b, the single-shot stack has sufficiently high SNR to resolve microscopic features, such as puncta, which are likely to be lipofuscin granules.
This is confirmed in the higher-SNR averaged image (Fig.~\ref{fig:zebrafish}a).
Lipofuscin granules are medically relevant -- their photoactivity can pose a danger to nearby cells, and a correlation has been observed between these granules and debilitating diseases such as advanced macular degeneration (AMD) \cite{feldman_photobiology_2022}.
Notably, both the single-shot and averaged lifetime images provide more information than intensity alone (Fig.~\ref{fig:zebrafish}d), which does not show variation in the epithelium lifetime, which could indicate increased oxidative stress \cite{feldman_photobiology_2022}.

In addition, the lifetime performance of the single-shot stack is comparable to the highly averaged stack.
The lifetime histogram in Fig.~\ref{fig:zebrafish}c shows that the mode of the single-shot lifetimes has a slight shift towards lower lifetimes, but still matches the 20$\times$-average distribution fairly well.
Quantitatively, the single-shot lifetimes differed from the 20$\times$ reference by a mean of -0.08\,ns.
This may be in part due to an error in calibration between the single-shot and averaged impulse response functions (the latter is $\sim$0.04\,ns broader due to averaging over timing jitter), as well as micron-scale drifts of the sample between the two acquisitions.
The intensity-weighted lifetime standard deviations across the analyzed region were 0.68 ns for single-shot acquisition and 0.53 ns for 20$\times$-averaging, corresponding to a modest reduction in spatial variability.
Lastly, Fig.~\ref{fig:zebrafish}f shows that the single-shot and averaged profiles of the mean lifetime with radial distance from the eye's center match quite closely.

%% file: text_kidney.tex
\subsection*{Deep, denoised lifetime measurement of a mouse kidney}\label{sec:kidney}
\addcontentsline{toc}{section}{\nameref{sec:kidney}}

\begin{figure}[!htbp]
\centering
\includegraphics[width=\normaltextwidth]{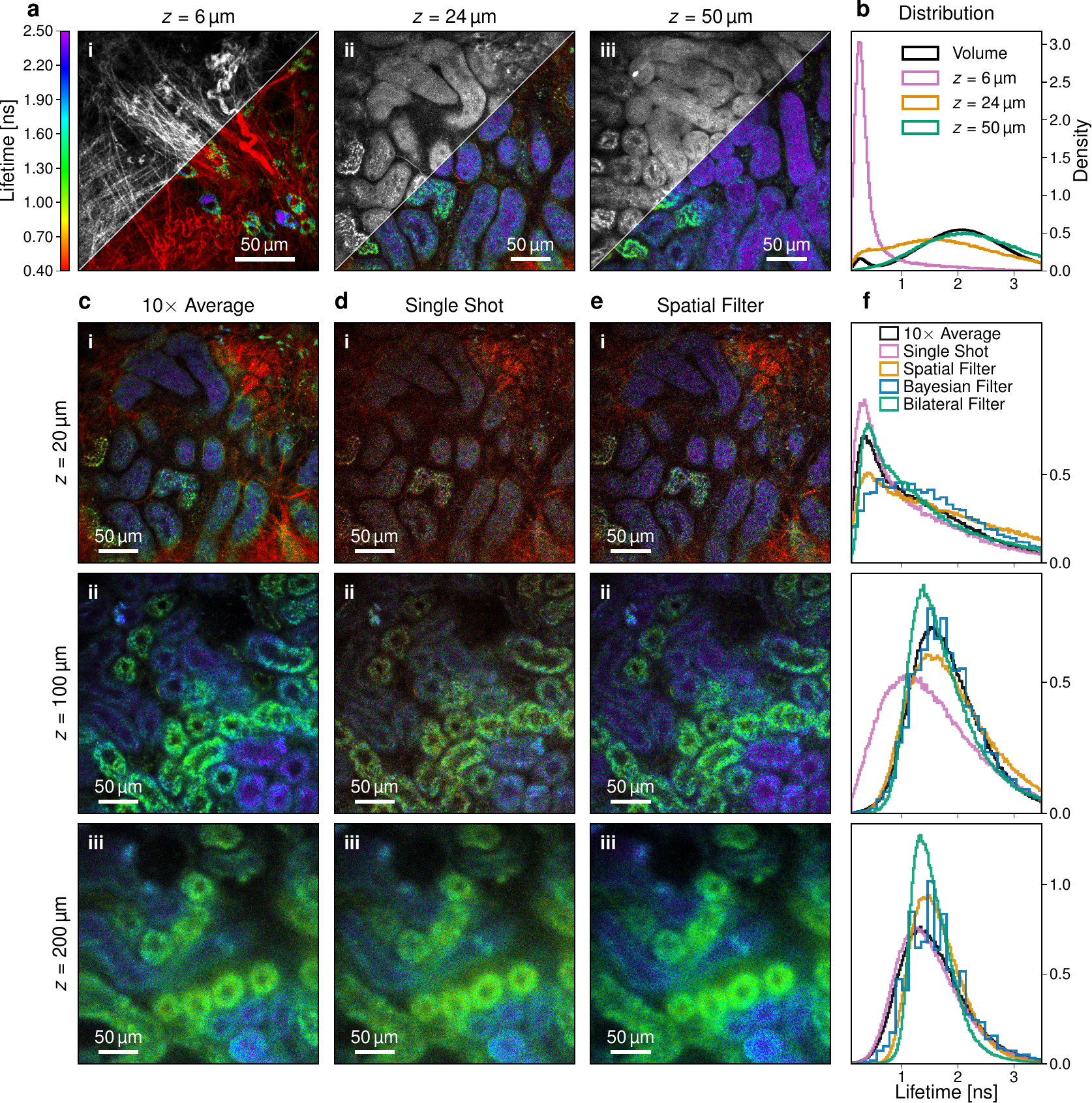}
\caption{\textbf{Deep lifetime imaging of a murine kidney.}
 \textbf{a.} Representative intensity (upper-left) and fluorescence lifetime (lower-right) images acquired at depths of $z=6$, 24, and 90 {\textmu}m.
 \textbf{b.} Intensity-weighted lifetime distributions corresponding to the three depths in \textbf{a}.
 \textbf{c--e.} Lifetime images taken at depths of $z=20$, 100, and 200\,{\textmu}m, respectively, using (\textbf{c}) a 10$\times$-averaged measurement, (\textbf{d}) single-shot fitting, and (\textbf{e}) the same single-shot data, where the temporal-decay data is passed through a spatial filter.
 \textbf{f.} Intensity-weighted lifetime distributions over the image slices for each estimator (single pixel and spatial, Bayesian and bilateral filters) compared with the and 10$\times$ averaged reference.
}
\label{fig:kidney}
\end{figure}

The kidney is a demanding and rewarding target for FLIM as metabolically distinct nephron segments are organized into intricate three-dimensional tubules buried beneath scattering, collagen-rich layers.
Lifetime contrast in these samples is harder to recover at depth but biologically informative \cite{hato_two-photon_2017, kwon_fluorescence_2026, goiffon_dynamic_2009, ranjit_phasor_2020}.
For example, Fig.~\ref{fig:kidney}a shows regions that have similar intensity but differing lifetimes.
In particular, it is much easier to distinguish cells on the lifetime-colored halves of Fig.~\ref{fig:kidney}a(i) than on the intensity halves.
Previously, two-photon FLIM of kidney samples has been demonstrated at depths around roughly 100\,{\textmu}m \cite{ryu_real-time_2018, ranjit_label-free_2016}.
Extending this depth is non-trivial because depth-related scattering limits fluorescence excitation and collection.
Nonetheless, we sought to extend this depth limit by combining low-repetition-rate excitation, single-shot full-waveform detection, and lightweight denoising.

To demonstrate this, we obtained two 400\,$\times$\,400\,$\times$\,200\,{\textmu}m lifetime stacks from a murine kidney -- using both 10$\times$ averaging and single-shot-per-pixel at each location.
The single shot volume is rendered in Fig~\ref{fig:intro}a.
The averaged image provides lifetime contrast suitable for distinguishing various tissue structures and organelles (Fig.~\ref{fig:intro}), including collagen, renal tubules, and bright puncta consistent with endosomes \cite{welling_sorting_2010}.
Corresponding slices from the averaged (Figs.~\ref{fig:kidney}c(i--v)) and single-shot (Figs.~\ref{fig:kidney}d(i--v) datasets show that the single-shot data reproduce the lifetime and intensity contrast observed in the averaged data.

However, despite providing similar contrast as the averaged dataset, it is clear from the images that the lifetime values in the single-shot dataset are noisier.
Furthermore, the lifetime histogram (Fig.~\ref{fig:kidney}f) reveals a broader lifetime distribution with a significant short-lifetime tail in the single-shot measurements.
These short-lifetimes are likely a consequence of the low-photon counts, where the lack of photons means that a full decay curve is not always realized.
At low photon counts and especially for longer lifetimes, fitting may be biased towards shorter lifetimes as seen in Fig.~\ref{fig:validation}e.
The significant and randomly distributed portion of pixels with an erroneously short lifetime tend to give the image as a whole a redder appearance, and the longer lifetime contrast ($\sim$2\,ns, purple) is less apparent as more individual pixels are prone to inaccuracy.
Paradoxically, this is more problematic at shallower depths; likely because the point spread function is broadened at depth and the tissue can thus sustain higher power and yield more signal (see \hyperref[supp:poweranddepth]{Supplementary}).
Compared to the eye tissue, the endogenous fluorescence in kidney tissue was relatively weak, and required significantly higher powers to observe meaningful lifetime contrast.
However, obtaining the same quality of images by means of greater excitation pulse energies was not possible due to the onset of photodamage, which we observed caused changes in the lifetime of the affected area prior to physically altering the tissue.

To address this intensity-dependent short-lifetime bias without resorting to alternative lifetime estimation methods to fitting, we considered denoising approaches that would operate directly on the digitized temporal waveforms.
The goal is to evaluate whether the single-shot images contain a comparable lifetime information content to higher-SNR, averaged counterparts.
An analysis on individual pixels may lead to a loss of lifetime information, but we expect that a global or wider-context analysis, which considers multiple traces and how they correlate, may be able to more faithfully reconstruct the underlying image.
We thus considered several methods of varying complexity: a spatial filter as a baseline;
a Bayesian approach, where a prior is computed from the neighborhood average;
and a bilateral filter, which besides proximity, correlates the similarity of the decay.
See \hyperref[supp:kidney]{Supplementary} for further results.
These approaches use information from neighboring pixels to improve the lifetime estimation, as this can provide context of the local fluorescence environment.
Indeed in both latter approaches -- and in most conceivable denoising approaches that leverage spatial information -- the goal is to pool signals that are deemed to be the same, while excluding nearby signals deemed to be separate species.
At the shot-noise limit, this spatial pooling effectively increases the number of photons available to estimate the local decay, by ``borrowing'' them from nearby pixels, restoring the measured decay waveform, hence providing a direct means to improve the SNR and precision.
All approaches can benefit by a change of basis to the frequency domain, as the change of basis compresses the data, and hence the processing time.
This can enable longitudinal, high-throughput biological studies without massive raw data acquisition and postprocessing.

The results show that all methods (Figs.~\ref{fig:kidney}c--f; \hyperref[supp:kidneycomparison]{Supplementary}) are able to correct the lifetime estimates, both qualitatively and by comparing distribution with respect to the averaged data.
In fact, the more sophisticated methods only provided minor improvement over the baseline spatial filter, which we thus show in Fig.~\ref{fig:kidney}e.
Note that the distributions do not capture the image fidelity and noise,
and in addition the averaged-image distributions are still broadened due to photon-number-limited lifetime uncertainty, and cannot be considered a true ground truth.
A direct comparison between single and averaged scans was not possible to due three-dimensional micron-scale shifts.

Altogether, these results show that single-shot lifetime acquisition can preserve recognizable intensity and lifetime contrast to depths of at least 200\,{\textmu}m in scattering kidney tissue, while the lifetime noise can be substantially reduced with lightweight denoising approaches that can operate in or close to real time.
Besides accuracy: complexity, generality and speed are important considerations for translational research and clinical application.
We expect that the results may be further improved with large data-driven models that jointly model the spatial and lifetime data \cite{zhou_high-fidelity_2026, kapsiani_deep_2025}.
While denoising lifetime estimation has been done in prior works, to the best of our knowledge it has not been applied to direct analog time sampling or related methods.

%% file: text_discussion.tex
\section*{Discussion}

Here we demonstrate that label-free fluorescence lifetime information can be recovered from as few as a single multiphoton excitation pulse in intact living biological systems.
By combining a low-noise detection front end whose read noise falls below one photoelectron with a jitter-immune, frequency-domain estimator, this approach achieves a photon-economy factor $F<1.5$ from single PMT traces containing as few as 100--150 detected photons, without averaging pulses or borrowing photons from neighboring pixels.
Complementary to photon-counting approaches, the proposed single-shot FLIM addresses regimes in which excitation rates must be kept low or acquisition volumes are large.
Now, lifetime measurements can be obtained from the same pulse budget used to form intensity images, enabled by the sole addition of a giga-sample digitizer, whose speeds, costs, and availability have been steadily improving.
This makes our proposed method particularly attractive for applications where low laser repetition rate and energetic pulses are preferred either for optimized dark-state relaxation \cite{willig_sted_2007} or deep-tissue imaging \cite{horton_vivo_2013}.
In addition, extending this method to multiple spectral channels is straightforward (see \hyperref[supp:multichannel]{Supplementary}), which would enable simpler pulse-sampled metabolic imaging and fluorescence lifetime multiplexing \cite{Hwang2024}.
Finally, improvements in photodetector technology will further increase performance and improve the fidelity of single-shot lifetime estimation: in the \hyperref[supp:hpd]{Supplementary}, we show that hybrid photodetectors can already achieve $F \leq 1.3$ in direct comparison to our PMT.

The zebrafish and kidney experiments demonstrate the capability of single-shot label-free FLIM in complementary biological regimes.
In the living, optically transparent zebrafish eye, single-shot acquisition reproduced the principal lifetime features of a 20$\times$ averaged reference.
In contrast, the intact, unsectioned kidney provided a more stringent test case because its collagen-rich capsule and densely scattering cortex caused severe signal loss.
In spite of this, recognizable tubular, cellular, collagen, and endosome-like lifetime contrast persisted across a 200 \textmu m deep volume.
Together, these measurements demonstrate that single- to few-pulse label-free volumetric FLIM is applicable both in living specimens and in deep, strongly scattering tissues.

Importantly, the kidney data also exemplifies the current boundary of the approach, at least at the single-shot limit that we have studied (in practice some averaging would be desirable).
Where photon counts were low, the measurements showed lifetime contrast on the whole, but exhibited noise and some negative bias for longer lifetimes, resulting in ``red speckle''.
We demonstrated that several denoising techniques are effective in mitigating this but are constrained by the limited decay information present in each photon-starved waveform.

There have been several recent works that integrate large deep-learning models and photon counting approaches.
They show encouraging evidence that, leveraging information from a broader spatial context or throughout successive frames can significantly boost the lifetime estimation fidelity with substantially fewer photons \cite{zhou_high-fidelity_2026, shen_overcoming_2024}.
We expect comparable gains can be achieved with the proposed pulse sampling FLIM platform.
However, the tradeoff is a significant one: fast imaging where the results can be viewed in real-time and requiring no training data; or, saving large amounts of raw data for postprocessing with a model trained for a specific instrument and possibly only for a given sample.
In addition, pooling across neighbors or frames risks spatial and temporal smearing, and overfitting to the training distribution unless an unsupervised method is used.
With future efforts in physics-informed, data-driven algorithms and rigorous testing protocols, we believe single-shot label-free FLIM can work below the demonstrated photon regime and serve as a major workhorse complementary to photon-counting approaches.

%% file: text_methods.tex
\section*{Methods}\label{sec:methods}
\addcontentsline{toc}{section}{\nameref{sec:methods}}

\subsection*{Multimode fiber}

The 1100\,nm pulses used for multiphoton imaging were generated by exploiting the self-broading in a 25\,{\textmu}m core multimode fiber (Thorlabs, FG025LJA).
The fiber was pumped by a 1030\,nm, 200\,fs mode-locked ytterbium laser (Light Conversion, Carbide-40W-10MHz), with a pulse energy of approximately 450\,nJ and a repetition rate of 1\,MHz.
The fiber was cut to a length of 3\,cm to match the length of the fiber groove (Thorlabs, HFV001) and to minimize the pulse temporal broadening.
The filtered pulse at the output of the fiber was measured as approximately 100\,fs in duration via autocorrelation.
A 50\,nm bandpass filter was used (Edmund Optics, 85-894).
After aligning the fiber, the power was fine tuned for the peak to coincide with 1100\,nm, as measured by a spectrometer (Ocean Optics, NIRQuest).
The beam after the fiber was collimated to approximately 8\,mm.
The filtered pulse energy was typically measured to be 80\,nJ.
This power was adjusted using a half-wave plate (EKSMA, 460-4208) mounted on a motorized rotation stage (Thorlabs, PRM1Z8) and polarizing beam splitter (Thorlabs, PBS253).

\subsection*{Imaging system}\label{methods:system}

\begin{figure}[htbp]
\centering
\includegraphics[width=0.75\normaltextwidth]{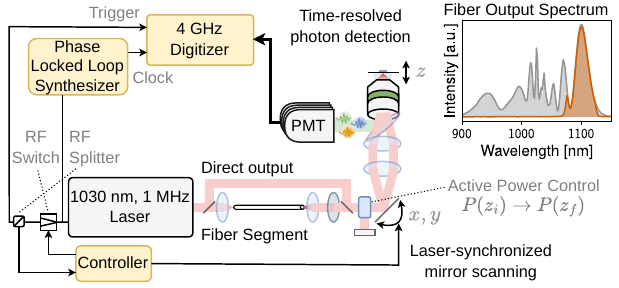}
\caption{
\textbf{Experimental Setup.}
}
\label{fig:setup_}
\end{figure}

See Fig.~\ref{fig:setup_} for a diagram.
The microscope was implemented in an inverted configuration.
A pair of scanning mirrors (Scannermax, Saturn 5B) was used to raster scan the sample by directing the excitation light through a water immersion objective (Olympus, XLPLN25XWMP2).
Two sets of lenses were used as an optical relay between the scanning mirrors and the objective (Thorlabs, one pair of AC508-075-C-ML and one each of AC254-030-C-ML, AC508-100-C-ML).
A microscope stage (Applied Scientific Instrumentation, MS-2000, LS-50-AMERL, LS-50-AMCLH) was used to move the sample into the focal plane and adjust the depth.
The excitation and emission signal beam paths were separated by a dichroic mirror (Semrock, FF925-Di01-25x36).
Light from the sample was collected (Thorlabs, LA1951-A) and detected by a photomultiplier tube (PMT) (Hamamatsu, H16722-40; powered by Siglent, SPD3303X).
The electronic signal from the PMT was sent to a digitizer (Alazar, ATS9373), installed in a PCIe slot of a desktop computer.
To synchronize the acquisition system, TTL pulses originating from the laser were used as the master clock.
These pulses were directed towards three other devices: the aforementioned digitizer, a phased-lock loop (PLL) frequency synthesizer (Analog Devices, EVAL-ADF4350EB1Z), and a function generator (Liquid Instruments, Moku Go).
The electronic pulses were used to trigger the digitizer.
In addition, the frequency synthesizer generated a clock signal for the digitizer, phase-locked to the laser trigger.
This minimized the time jitter between the signal arrival time and the start of each trace acquisition.
Finally, the function generator had the responsibility of 1) incrementing the scanning mirror voltage with each laser trigger event, and 2) opening and closing the path between the laser trigger to the digitizer at the beginning and end of each acquisition via a radio-frequency switch (Mini-circuits, ZFSW-2-46).
This was in order to 1) guarantee that the mirror angles at the exact time of a laser pulse coincided between frames, and 2) that the traces acquired by the digitizer could be deterministically assigned to pixels.
All digital signal processing and instrument control for the microscope was done via custom software written in Python.


\subsection*{Volumetric imaging}

3D volume images were generated by stacking 2D images acquired at various depths, acquired by moving the sample with respect to the focal plane using the microscope stage.
The pulse energy was adjusted at each layer, by programmatically rotating the half wave plate described previously. 
Pulse energy in the focal plane ranged from 0.5 to 1.5 nJ for the kidney, and was around 0.5 nJ for the eye imaging.
This led to a maximum number of observed photons of around 30 for the kidney, and 50 for the brighter and less scattering eye.
The resulting detected-photon estimates for in vivo and ex vivo imaging are of a similar order of magnitude to photon estimates reported for multiphoton microscopy in the literature \cite{Wang2024Pixelation}.
Averaging was done by repeating the 2D scan before proceeding to the next layer.

The eye and dyes were imaged using the filtered 1100\,nm fiber output (see above), while the kidney was imaged using 1030\,nm pulses directly from the laser.

Visualizations were prepared using the Napari 3D viewer, with additive blending and attenuated maximum intensity projection rendering.
An exponential depth-dependent intensity gain $g = \exp(\alpha z)$ was applied, where $z$ is the layer and constant $\alpha=2.5$.

\subsection*{Fluorescence lifetime estimation}\label{methods:fitting}

Traces containing decay curves were analyzed on the GPU.
A custom-written Levenberg--Marquardt algorithm was used.
The detailed procedure is as follows.
The analog-to-digital acquisition was configured to include a short delay between the beginning of the acquisition window and the arrival of the emission signal.
The mean value of this pre-signal window was averaged over the whole scan row, and then subtracted from each trace.
This was necessary to remove the DC offset, which was subject to a slow drift (microsecond timescales).
The total acquisition time was chosen to be 24\,ns, or 96 time bins at 4\,Gs/s (this was kept fixed in the scope of this work, but can otherwise be truncated or zero-padded).
This obeys the rule of thumb that the acquisition window must be at least 4 times longer than any lifetime measured \cite{dowell_precision_1988}.
Once centered, batched real fast Fourier transform (rFFT) was performed on the traces.
In order to negate the effects of time shifts and jitter, the complex phase of the frequency-domain signal was removed by taking the absolute value (modulus) of the spectra.
By removing the time-shift degree-of-freedom, only the amplitude and lifetime of the signal remain to be jointly fitted.
Furthermore, the frequency-domain signal was truncated, down to a bandwidth of 0.5\,GHz, corresponding to the bandwidth of the PMT impulse response (see Fig.~\ref{fig:validation}b).
This excluded bins containing predominantly or only noise, and corresponds to 12 frequency bins.
This cuts down the number of data points involved in nonlinear least squares fit by a factor of 8.
The resulting frequency response data, after rFFT, modulus, and truncation, represents the target function for the fit.
The fit model $f$ is the convolution of the instrument impulse response (IR) with a Lorentzian-like function:
\begin{equation}
f(\omega; a, \tau) = H(\omega) \: \frac{a}{\sqrt{1 + (\omega \tau)^2}}
\end{equation}
Here $\omega$ is the angular frequency, the independent variable in the fit, $a$ and $\tau$ are respectively the amplitude and lifetime, the dependent variables, and $H$ is the frequency IR (real-valued).
By fitting in Fourier space we avoid the back-and-forth transforms necessary for the convolution operation between signal and IR.
In addition, the problem reduces to estimating the width of a curve via a monotonically decreasing function, which greatly simplifies the optimization landscape compared to fitting a pulse, which translates to faster convergence.
During fitting, the Levenberg--Marquardt algorithm is run in a batched fashion for a fixed number of iterations, specifically 50, although the vast majority of fits typically converged within 10--20 iterations.
The initial parameters were estimated using a heuristic based on the spectral centroid.
The fitting program was implemented in Torch for ease of development, custom-written specifically to fit the above function.
Including data transfers to and from main memory and GPU, the average computation time per trace asymptoted to approximately 0.35\,{\textmu}s for 50 iterations on an NVIDIA GeForce RTX 3080 Ti GPU.
This performance could be further improved for higher repetition rates by direct memory access for streaming GPU memory, which is supported by these digitizers, by implementing the same optimization loop directly in a CUDA kernel, at the expense of portability, by reducing the number of data points or iterations, as well as by using a higher-performance GPU.

\subsection*{Denoising}\label{methods:denoising}

\subsubsection*{Spatial Filter}
The spatial filter consisted of a Gaussian kernel with parameter $\sigma_s = 0.5$ pixels, or 0.25\,{\textmu}m, about half the diffraction-limited excitation focal diameter.
The filter was applied after FFT, on the truncated complex-valued spectrum, and the result was passed to the fitter.

\subsubsection*{Bilateral Filter}
Bilateral filters are used for denoising while preserving edges and sharp features.
A Gaussian kernel was used for the spatial component (pixel distance), with a broader width ($\sigma_s = 2$ pixels) in order to pool from a larger area (greater number of pixels) compared to the naive spatial filter.
The kernel pooled within a $\sqrt{17}$ pixel radius.
A Gaussian was also used for the \emph{range} kernel:
to perform discriminate spatial averaging, by only considering similar traces.
The range weight was assigned based on the similarity of the nearby traces according to the Euclidean distance of the truncated complex Fourier spectra as described above.
The Gaussian width used was $\sigma_r = 10$\,mV\,ns.
This prevented pooling traces with either different intensities or lifetimes.
Overall, the filter spectra $\hat f(\omega)$ was computed from the original spectra $f(\omega)$ as follows:
\begin{equation}
\hat f(\omega; x, y) = \sum_{x', y'} \exp\!\left( -\frac{x^2 + y^2}{2\sigma_s^2} -\frac{|f(\omega; x, y) - f(\omega; x', y')|_\omega^2}{2\sigma_r^2} \right) f(\omega; x', y')
\end{equation}
where $x,y$ are the image pixel coordinates, and $x',y'$ are the neighboring pixel coordinates.

\subsubsection*{Bayesian Denoising}

To incorporate spatial information into lifetime estimation, we construct a local prior distribution for the fluorescence lifetime at each pixel or voxel. 
For each location, we first calculate the average temporal trace of its spatial neighborhood, excluding the trace at the location being estimated. 
The neighborhood is defined as a \(3\times3\) window for two-dimensional images or a \(3\times3\times3\) window for three-dimensional volumes, with reflective boundary conditions. 
We then fit the averaged neighborhood trace using the same lifetime-fitting procedure used to initialize the lifetime estimates, obtaining a local neighborhood lifetime \(\tau_{\text{neigh}}\).
The prior probability distribution over candidate lifetimes \(\tau\) is modeled as a Gaussian centered at \(\tau_{\text{neigh}}\):
$$
P(\tau) \propto
\exp\left[-\frac{(\tau-\tau_{\text{neigh}})^2}{2\sigma_{\text{spatial}}^2}\right],
$$
where \(\sigma_{\text{spatial}}\) controls the strength of the spatial prior.
The distribution is evaluated over the discrete lifetime grid and normalized to sum to unity, with a small additive constant to prevent zero probabilities. 
This prior favors lifetimes consistent with the surrounding neighborhood while allowing deviations when supported by the measured temporal trace.

For each candidate lifetime, we evaluate the likelihood of the measured temporal trace using a Monte Carlo model of the detector signal.
For a candidate lifetime $\tau$ and estimated detected photoelectron count $K$, we generate $M$ simulated realizations of photon-arrival times by sampling from an exponential decay distribution with lifetime $\tau$. 
The model also incorporates temporal jitter and a small count-dependent delay term.
Each simulated arrival is convolved with a Gaussian single-photoelectron response, and the responses are summed to generate a predicted detector trace.
The likelihood of the measured trace is evaluated assuming additive, independent Gaussian readout noise with standard deviation $\sigma_{\text{readout}}$.
The likelihood marginalized over the simulated arrival-time realizations is
$$
P(Y\mid\tau,K)\approx
\frac{1}{M}\sum_{m=1}^{M}
P(Y\mid\tau,K,m),
$$
where $Y$ is the measured temporal trace and $m$ indexes the Monte Carlo realizations.
A constant baseline is estimated from the initial samples of the measured trace and removed before calculating the residuals between measured and simulated traces.

We combine the likelihood with the spatial prior using Bayes’ theorem:
$$
P(\tau\mid Y,K)=
\frac{P(Y\mid\tau,K)P(\tau)}
{\sum_{\tau'}P(Y\mid\tau',K)P(\tau')},
$$
where the denominator normalizes the posterior over the candidate lifetime grid.
Calculations are performed in the logarithmic domain for numerical stability.
The final lifetime estimate is calculated as the posterior-weighted mean of the candidate lifetimes,
$$
\hat{\tau}=\sum_{\tau}\tau \cdot P(\tau\mid Y,K),
$$
where the sum is taken over the discrete lifetime grid.
This estimator incorporates both the temporal evidence from the measured trace and the spatial information from neighboring locations, favoring spatially consistent estimates when the temporal data are noisy while allowing deviations when supported by the measured signal.

\subsection*{Photomultiplier tube signal photon number estimation}\label{methods:penumber}

In order to estimate the number of photons incident onto the PMT, we integrate the trace (after subtracting the DC bias, as described above).
By appropriately scaling the analog-to-digital conversion (ADC) units to physical units, this yields a value in volt nanoseconds (V\,ns), hence an electrical charge after accounting for the 50\,$\Omega$ transmission line impedance.
We divide this value of total charge by the electron charge and the PMT gain, which was provided in the datasheet (in all relevant experiments we use a fixed control voltage of 0.7\,V, which equates to a gain of $4 \times 10^5$).
This value was independently validated by measuring the PMT output signal distribution, the results are presented in the \hyperref[supp:pmtpe]{Supplementary}.

The resulting quantity is the estimated number of photoelectrons generated by the PMT.
Since the PMT exhibits gain variation, each initial photoelectron produces a variable amount of current, so this is an estimate rather than a precise quantity, but it is correct on average.
As the number of photoelectrons relates directly to the number of captured photons (up to the quantum efficiency), the estimated incident photon number $\hat n$ is given by:
\begin{equation}
\hat n = \frac{\Delta t}{Z g e^-} \sum_i v[i] ,
\end{equation}
where $\Delta t$, $v[i]$, $Z$, $g$, $e^-$ are the ADC time step, ADC voltage values comprising the trace, impedance, gain and electron charge, respectively.
Transmission losses should also be accounted for but these were found to be negligible if the PMT was connected directly to the digitizer via a coaxial cable.

\subsection*{Sample Preparation}
In Fig.~\ref{fig:validation}, fluorescence lifetime calibration solutions were prepared using Rose Bengal (Thermofisher Scientific, 419150250), Rhodamine B (Thermofisher Scientific, A13572.30), and Fluorescein (Thermofisher Scientific, L13251.22).
For Rose Bengal and Rhodamine B, a stock solution was prepared by mixing 2\,mg of dye with 5\,mL of phosphate buffered saline (PBS) solution (Thermofisher Scientific, 14190144).
Prior to imaging, a dilution was made using 5\,mL of stock solution and 185\,mL of PBS.
For Fluorescein, a stock solution was made by mixing 2\,mg of dye with 5\,mL of 70\% ethanol.
Prior to imaging, a dilution was made using 20\,mL of stock solution and 180\,mL of 70\% ethanol.
All solutions were imaged in a glass bottom petri dish (Cellvis, D35-20-1-N).

In Fig.~\ref{fig:zebrafish}, Zebrafish lines were housed in Association for Assessment and Accreditation of Laboratory Animal Care (AAALAC) accredited facilities and maintained in accordance with protocols approved by the Massachusetts Institute of Technology Committee on Animal Care.
For label-free imaging, embryos were collected at the one-cell stage and incubated at 28°C until 24--36 hours post-fertilization.
Embryos were then anesthetized with tricaine and embedded in low-melting-point agarose on glass coverslips to immobilize them for subsequent imaging.

In Fig.~\ref{fig:kidney}, 12--16 week-old adult C57BL/6 mice were sacrificed by CO2 euthanasia and their kidneys were harvested.
Mice were housed in a 12 hour light-dark cycle with ad libitum access to water and standard laboratory chow.
All animal procedures were conducted in accordance with protocols 2021N000252 approved by the Institutional Animal Care and Use Committee (IACUC) at Mass General and conducted in accordance with the regulations of AAALAC International.

%% file: text_supplementary.tex
\begin{appendices}

\setcounter{figure}{0}
\setcounter{table}{0}
\renewcommand{\thefigure}{S\arabic{figure}}
\renewcommand{\theHfigure}{S\arabic{figure}}
\renewcommand{\thetable}{S\arabic{table}}
\numberwithin{equation}{section}

\renewcommand\appendixname{}

\vspace{\baselineskip}
{\centering \noindent\rule{0.5\normaltextwidth}{0.4pt}\par}
\vspace{\baselineskip}

\startcontents[sections] 
{
\hypersetup{linkcolor=black}
\section*{Supplementary Materials}
\printcontents[sections]{}{0}{}
}

\vspace{\baselineskip}
{\centering \noindent\rule{0.5\normaltextwidth}{0.4pt}\par}
\vspace{\baselineskip}

\section{Analysis of the PMT Response Waveform}\label{supp:analysis}

\subsection{Comparison of Analysis Methods}\label{supp:estimation}

\begin{figure}[!htbp]
\centering
\includegraphics[width=\normaltextwidth]{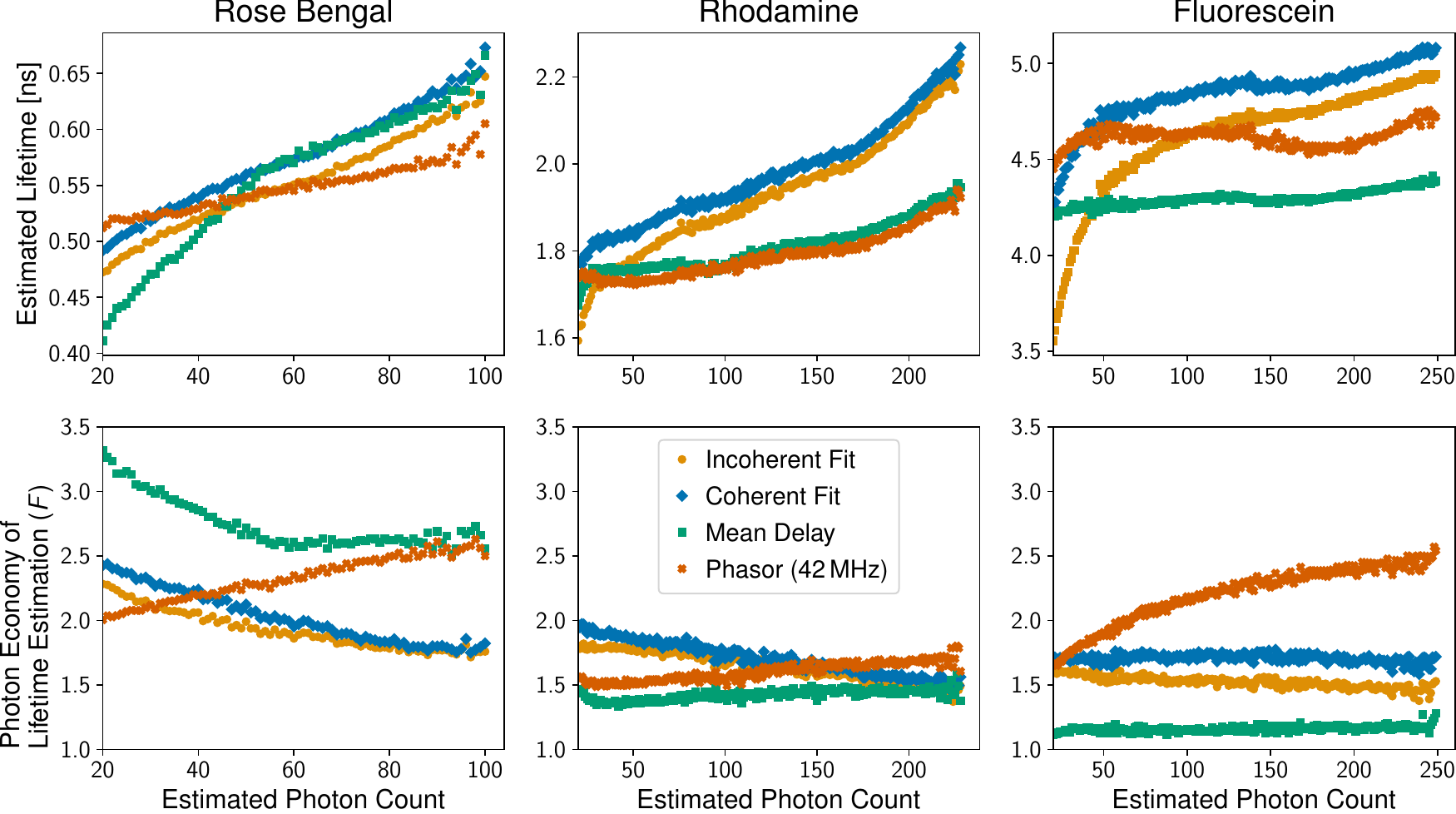}
\caption{
\textbf{Comparison of different lifetime estimation methods.}
We expand on the results in Fig.~\ref{fig:validation} by comparing other analysis methods.
}
\label{fig:comparisons}
\end{figure}

Using the same data from Fig.~\ref{fig:validation}, we compare our frequency amplitude fitting method (``incoherent'') with other methods that are used in lifetime imaging.
``Coherent'' fitting -- over the complex value spectrum with phase and additional fit parameter $\delta t$, the delay; the mean delay method, and the phasor method (using the phase delay of first non-zero frequency bin).

Comparing different lifetime analysis methods reveals the advantages and disadvantages of each technique.
In summary, time domain methods tend to perform comparatively well for long decay times, where the signal is distributed over the time domain, where jitter or noise have a relatively minor effect on the structure of the response.
However performance drops dramatically for short lifetimes, where the signal is highly localized in time and thus more susceptible to these effects.
The contrary is true for frequency domain analyses, which perform comparatively better for short lifetimes and worse for longer lifetimes.
For our system using PMTs we observe the crossover to occur at approximately 2\,ns.

While fitting is yields the most consistent photon economy, mean comes close to optimal at long lifetimes, as has been reported in Ref.~\cite{moon_analog_2009}, but suffers at short lifetimes, providing the worst performance.
The phasor method -- which relies on the complex phase of the first non-zero frequency bin -- has ostensibly good performance at low photon numbers, but scales poorly with respect to to photon number, and the photon economy increases rather than decrease compared to the other methods which improve, presumably due to improving with signal-to-noise.
An optimal analysis might combine various methods, weighing the various outcomes based on the measurement, to exploit each method in the lifetime--intensity regime in which it performs the best.

\begin{figure}[!htbp]
\centering
\includegraphics[width=0.8\normaltextwidth]{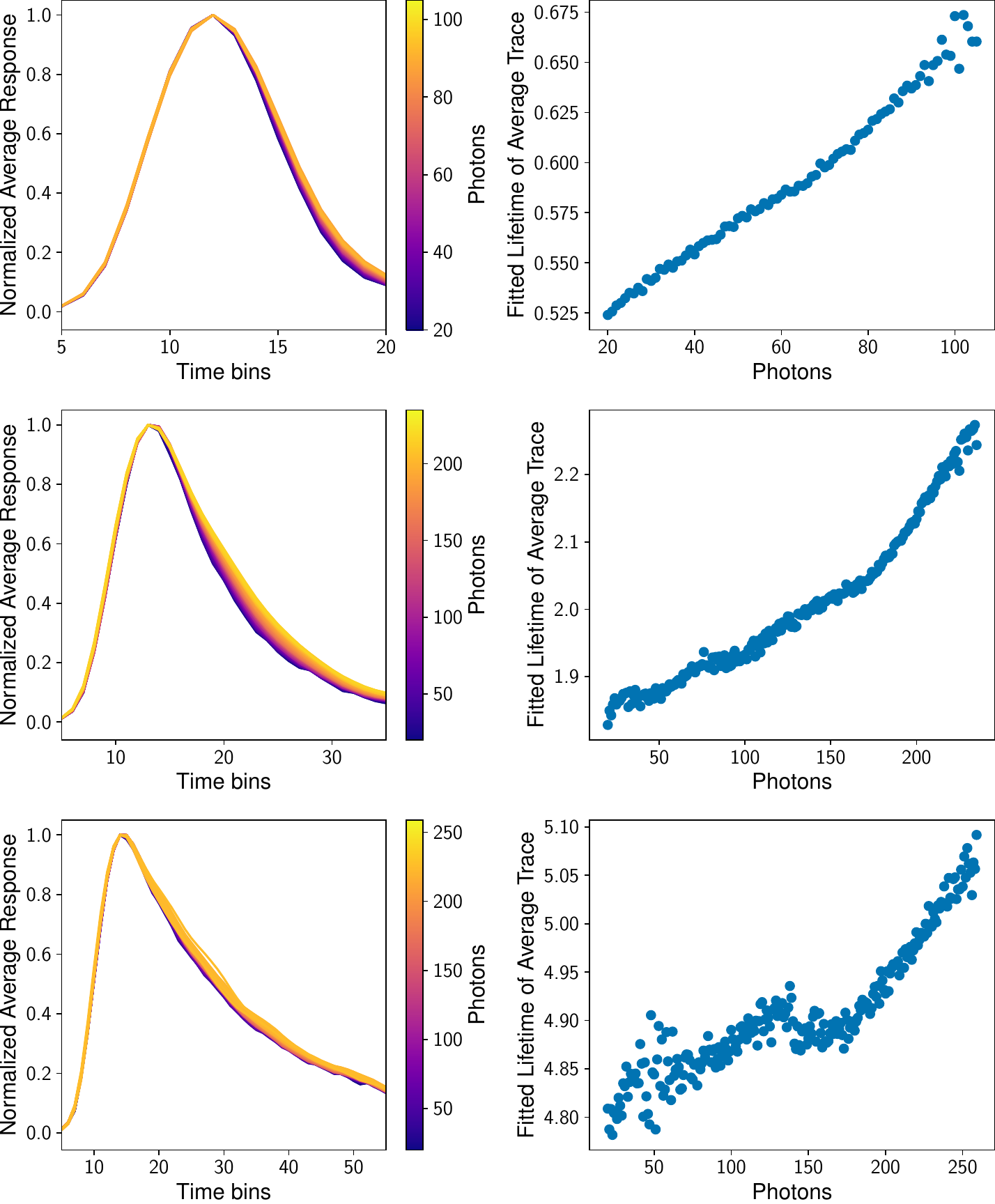}
\caption{
\textbf{Change in waveform with intensity.}
After binning traces by estimated photon number, superimposing plots of these traces show the tendency of the responses tend to smoothly broaden with intensity.
}
\label{fig:intensitydependence}
\end{figure}

For a variety of reasons, the different methods we consider here return marginally different lifetimes for the same data.
One of the major contributors to this is the pulse broadening observed in the PMT (Fig.~\ref{fig:intensitydependence}), which affects the tail of the exponential decay, while the rise remains identical.
This nonlinearity is likely due to space charge effects, and could in principle be calibrated based on the response intensity and decay profile.

\subsection{Change in Waveform with Intensity}\label{supp:nonlinearity}

Here we average the traces of rose bengal, rhodamine and fluorescein from the same intensity bin (Fig.~\ref{fig:validation}).
Plotting the normalized average traces reveals the broadening of the trace with photon number.
This broadening results in an increase in the fitted lifetime value:
the rise time remains largely unchanged, but the decay curve is broadened.
We do not model and characterize the nonlinearity in this work, which is likely due to space--charge effects specific to PMTs, and for which different photodetectors may have different nonlinearities.
However, we note that calibration and compensation of these kinds of effects must be developed in order for pulse sampling to validate as a standard method.

\subsection{Photomultiplier Tube Photon Number Validation}\label{supp:pmtpe}

\begin{figure}[!htbp]
\centering
\includegraphics[width=.58\normaltextwidth]{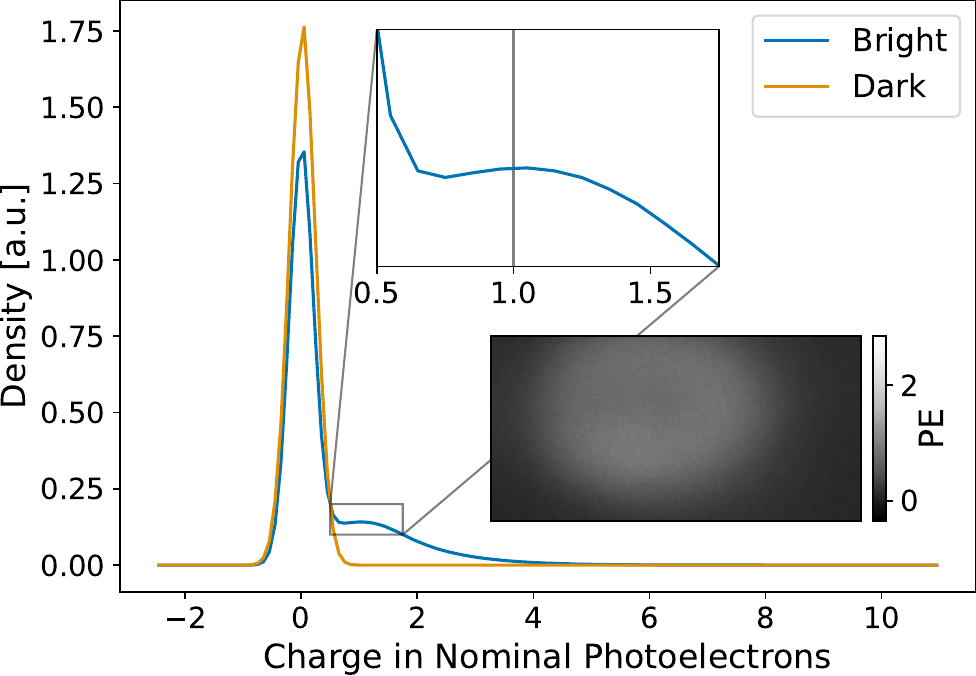}
\caption{
\textbf{Photomultiplier tube single photoelectron (PE) charge validation.}
}
\label{fig:photonnumber}
\end{figure}

As discussed in the methods, the photomultiplier tube has specified gains for given applied voltage.
By measuring low intensity third-harmonic generation signals (100 images of the same field of view), we validate that the gain closely matches expectation, in order to find the single photoelectron (SPE) charge.
This data was obtained by taking a set of pictures at a water--glass interface, and a histogram is shown in Fig.~\ref{fig:photonnumber}.
The histogram shows a clear peak very close to the nominal value of 1 photoelectron.
Nonetheless, we note that there is still uncertainty as to the mean SPE charge.
For example, the mode may be shifted due to contributions from higher photon numbers, or the well-documented skew of the SPE distribution may shift the average with respect to the mode. 
The resulting detected-photon estimates for biological tissue imaging are of the same order of magnitude to photon estimates reported for multiphoton microscopy in the literature \cite{Wang2024Pixelation}.

\subsection{Photon Economy of Hybrid Photodetectors}\label{supp:hpd}

\begin{figure}[!htbp]
\centering
\includegraphics[width=\normaltextwidth]{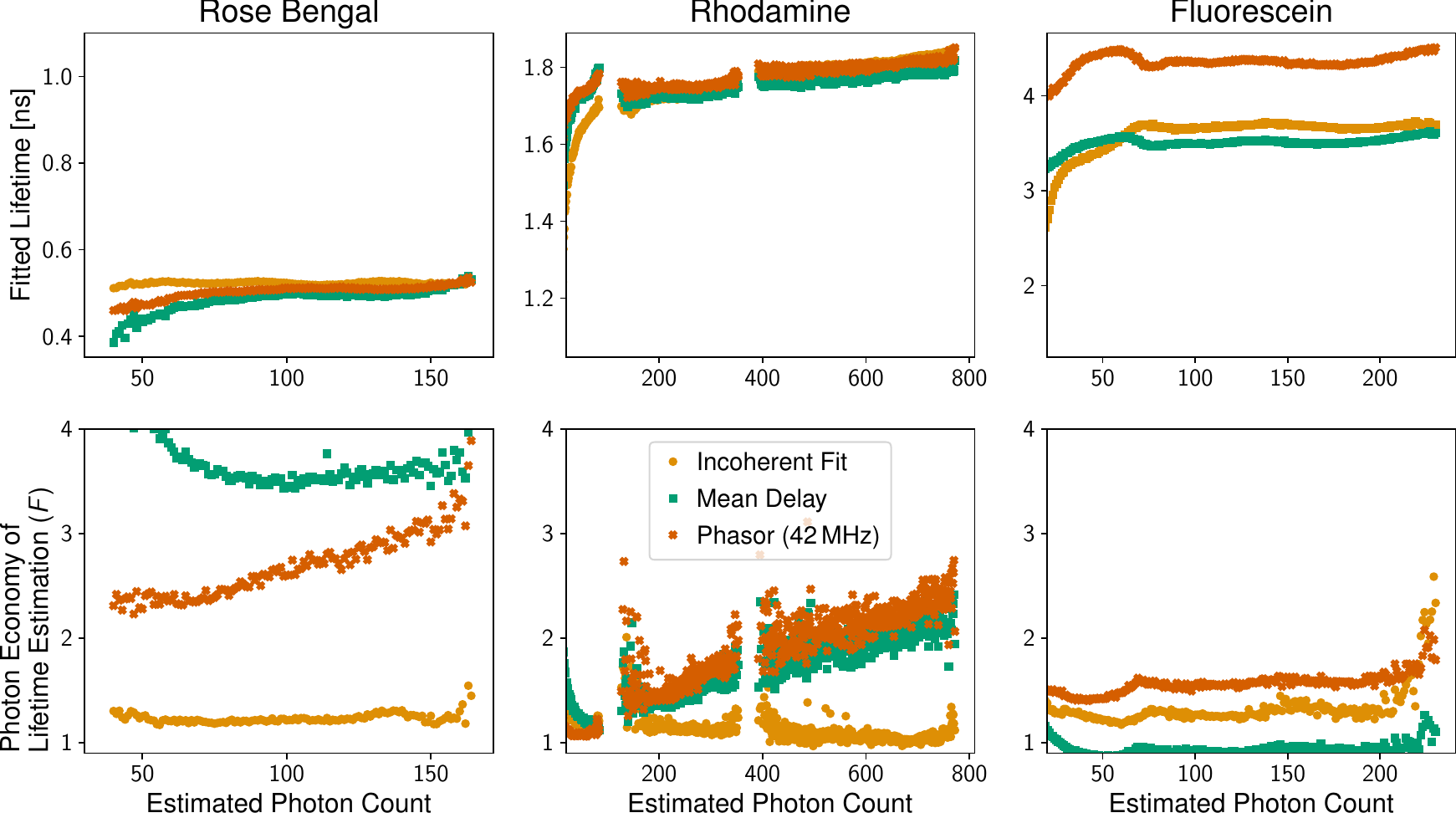}
\caption{
\textbf{Photon economy of hybrid photodetectors.}
The same analysis as Fig.~\ref{fig:validation} is shown, on data captured with a hybrid photodetector.
}
\label{fig:hpd}
\end{figure}

HPDs are modern photodetectors with low noise characteristics and with a response time that is approximately twice that of PMTs.
Despite being higher-fidelity hardware, we found that the absolute-time methods have worse performance than with PMT traces (see above), as timing noise is less forgiving with a narrower impulse response function.
Most likely this is due to time jitter or triggering inconsistencies.
In these cases, fitting drastically improves the photon economy until the lifetime exceeds that of Rhodamine, hence at least 2\,ns.

\subsection{Convergence of the Fitting with Iteration Number}

\begin{figure}[!htbp]
\centering
\includegraphics[width=\normaltextwidth]{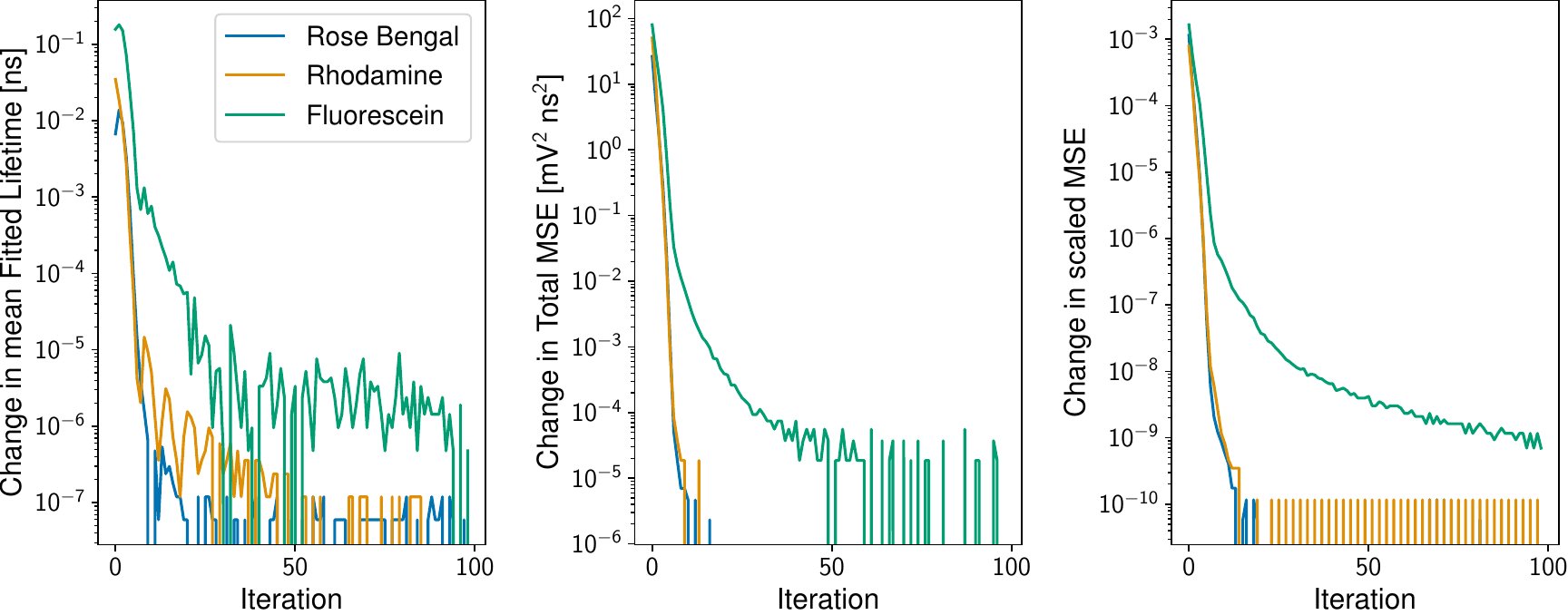}
\caption{
\textbf{Convergence of the fitting with iteration number.}
 We plot the results at each step during the iterative process of fitting the Rose Bengal, rhodamine and fluorescein datasets in Fig.~\ref{fig:validation}.
 Shown: the change in the lifetime between iteration steps;
 change in the mean-squared error (MSE) for each dataset, between iterations, given in physical units of (mV\,ns)$^2$;
 change in the scaled MSE, between iterations, defined as the MSE divided by the square of the trace integral.
}
\label{fig:convergence}
\end{figure}

Here we show the convergence of the custom fitting algorithm.
Namely, we show how 20 iterations or fewer are typically needed to converge to the minimum.
Although GPUs libraries are are becoming more ubiquitous, most optimization tools are optimized for deep learning, and do not second-order optimizers, that tend not to approach but never converge to the minimum.
These are thus impractical to use, especially for real-time imaging.
As 32-bit floating point numbers were used for GPU calculations, we expect that results relying on differences close to or past the 7 decimal digits of precision are dominated by numerical truncation error.
The runtime was approximately 0.65\,{\textmu}s for 100 iterations, and 0.35\,{\textmu}s per trace for 50 iterations, as typically used during imaging.

\clearpage
\section{Kidney Imaging}\label{supp:kidney}

\subsection{Spatial Filtering Baseline and Improved Methods}\label{supp:kidneycomparison}

\begin{figure}[!htbp]
\centering
\includegraphics[width=1\textwidth]{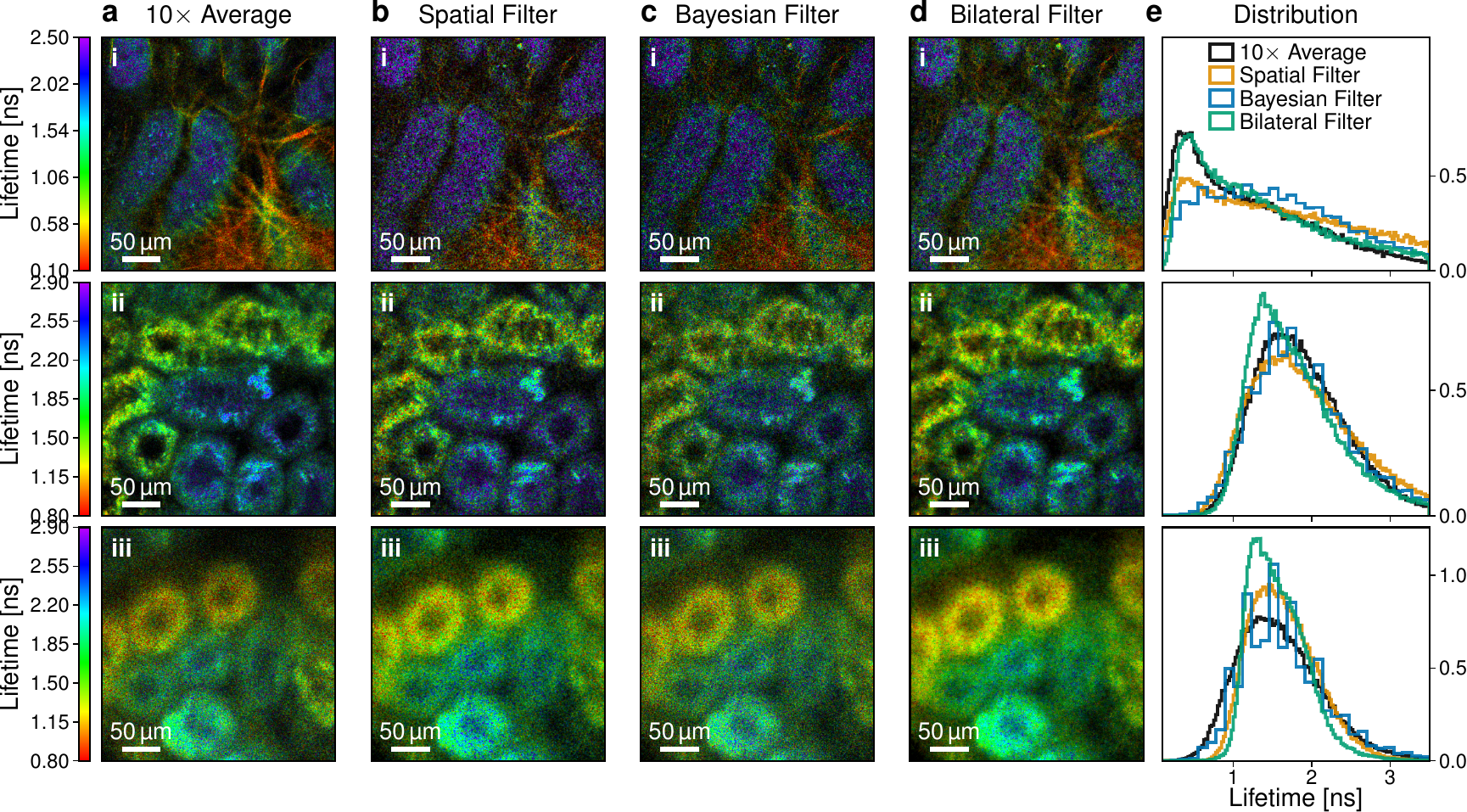}
\caption{\textbf{Comparison of different spatial filters.}
 Following Fig.~\ref{fig:kidney}:
 \textbf{a--d.} Lifetime images reconstructed from (\textbf{a}) a 10$\times$-averaged and single excitation pulses at depths of $z=20$, 80, and 200\,{\textmu}m, respectively, using (\textbf{b}) single-pixel fitting, (\textbf{cd}) spatial Bayesian inference, and (\textbf{d}) bilateral filtered.
 Note the use of different lifetime color maps.
 \textbf{e.} Intensity-weighted lifetime distributions over the image regions.
}
\label{fig:kidneysuppl}
\end{figure}

A common smoothing or denoising method is spatial filtering, however, this comes at the cost of potentially blurring the image if the filter has greater width than the imaging resolution.
We used a spatial filter as a baseline method in Fig.~\ref{fig:kidney}, and here in Fig.~\ref{fig:kidneysuppl} we include images processed via more advanced methods, over smaller regions of interest to show detail, noise and resolution.

The spatial filter consisted of a Gaussian with a $\sigma=0.5$ pixels, or 0.25\,{\textmu}m, about half the diffraction-limited excitation focal diameter.
We found that even this small resolution filter can compensate for low-photon-number lifetime noise, although there is a slight bias towards longer lifetimes.
Since the signal was sufficiently high, this sub-diffraction-limit-level averaging provided good performance without introducing significant blurring, although this will not always be the case.

The Bayesian approach involved taking the unweighted average trace of all surrounding pixels -- $3 \times 3 \times 3$ volume -- and using this pooled trace to construct a prior distribution.
A posterior, reconstructed trace, was found by combining this prior with the measured trace.
This implementation approximates the averaged distribution more closely when the lifetime contrast is spatially smooth.
When the neighborhood is highly heterogeneous, such as the thin collagen bundles found in Fig.~\ref{fig:kidneysuppl}(i) the distribution may fail to capture these lifetimes.

Bilateral filters are a classical image denoising method that preserve edges and sharp features, by performing discriminate spatial averaging.
This implementation involved a weighted sum of traces based on both pixel distance, and the difference between neighboring traces with respect to the center trace of interest.
Specifically the filter consisted of a Gaussian in both image space ($\sigma = 2$ pixels) and Euclidean distance in trace-Fourier space ($\sigma=10$\,mV\,ns), pooling within a $\sqrt{17}$ pixel radius.
Weighting by trace-distance avoids pooling traces with different intensities and lifetimes.
Indeed Fig.~\ref{fig:kidneysuppl} shows that the graininess is substantially reduced without a noticeable impact on resolution.
By being able to pool from a larger area (number of pixels), the method is also able to narrow the distribution beyond the 10$\times$-averaged: for example in Fig.~\ref{fig:kidneysuppl}(iii), the pixel distribution below 1\,ns is suppressed, as the lifetime image is ostensibly bimodal around 1.2\,ns and 2\,ns.

Note again that the averaged images are slightly shifted with respect to the processed single shot images, including along the $z$ axis.

\subsection{Power versus Depth in Kidney Imaging}\label{supp:poweranddepth}

\begin{figure}[!htbp]
\centering
\includegraphics[width=0.5\normaltextwidth]{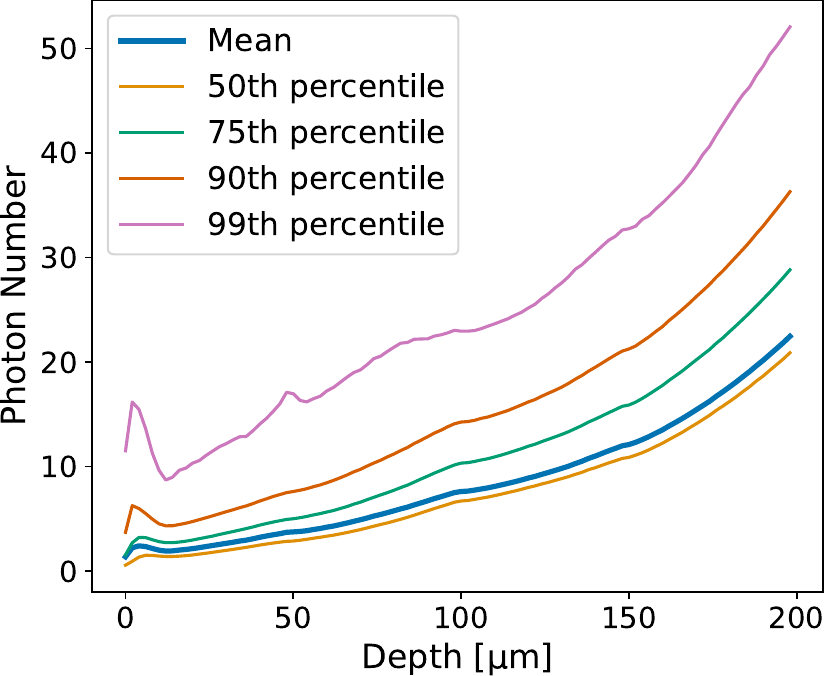}
\caption{
\textbf{Signal intensity with depth in kidney imaging.}
  For each image taken at a given depth, the intensity percentiles are plotted in units of photon number.
}
\label{fig:kidneyphotons}
\end{figure}

As discussed with regard to Fig.~\ref{fig:kidney}, the lifetime estimation tends to be less consistent at shallower layers compared to deeper layers.
We hypothesized that this is because in deeper layers, after propagating through the tissue, the point-spread function of the focused laser is broadened.
Since the focus is more spread out, greater power can be used for excitation without burning and a greater volume is excited, therefore yielding higher signals.
Fig.~\ref{fig:kidneyphotons} corroborates this idea, where we plot statistics of the intensity, in photon number.
The signal intensity increases with depth.
At shallow layers the risk of burning the sample due to the comparatively tighter focal spot (point-spread function) forced the use of a lower power, yielding fewer photons.

\subsection{Lifetime Histograms with Depth in Kidney Imaging}

\begin{figure}[!htbp]
\centering
\includegraphics[width=0.9\normaltextwidth]{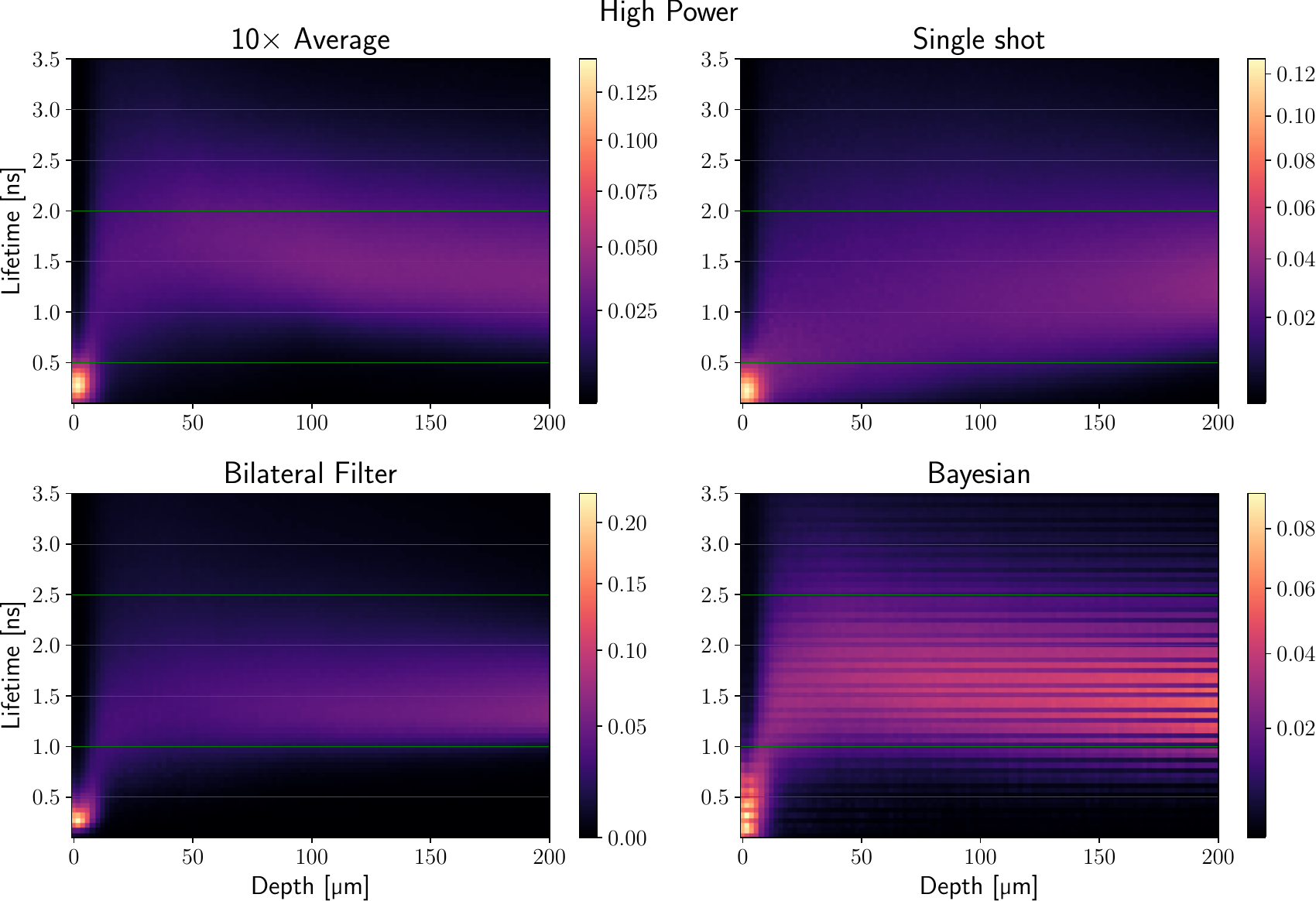}
\includegraphics[width=0.9\normaltextwidth]{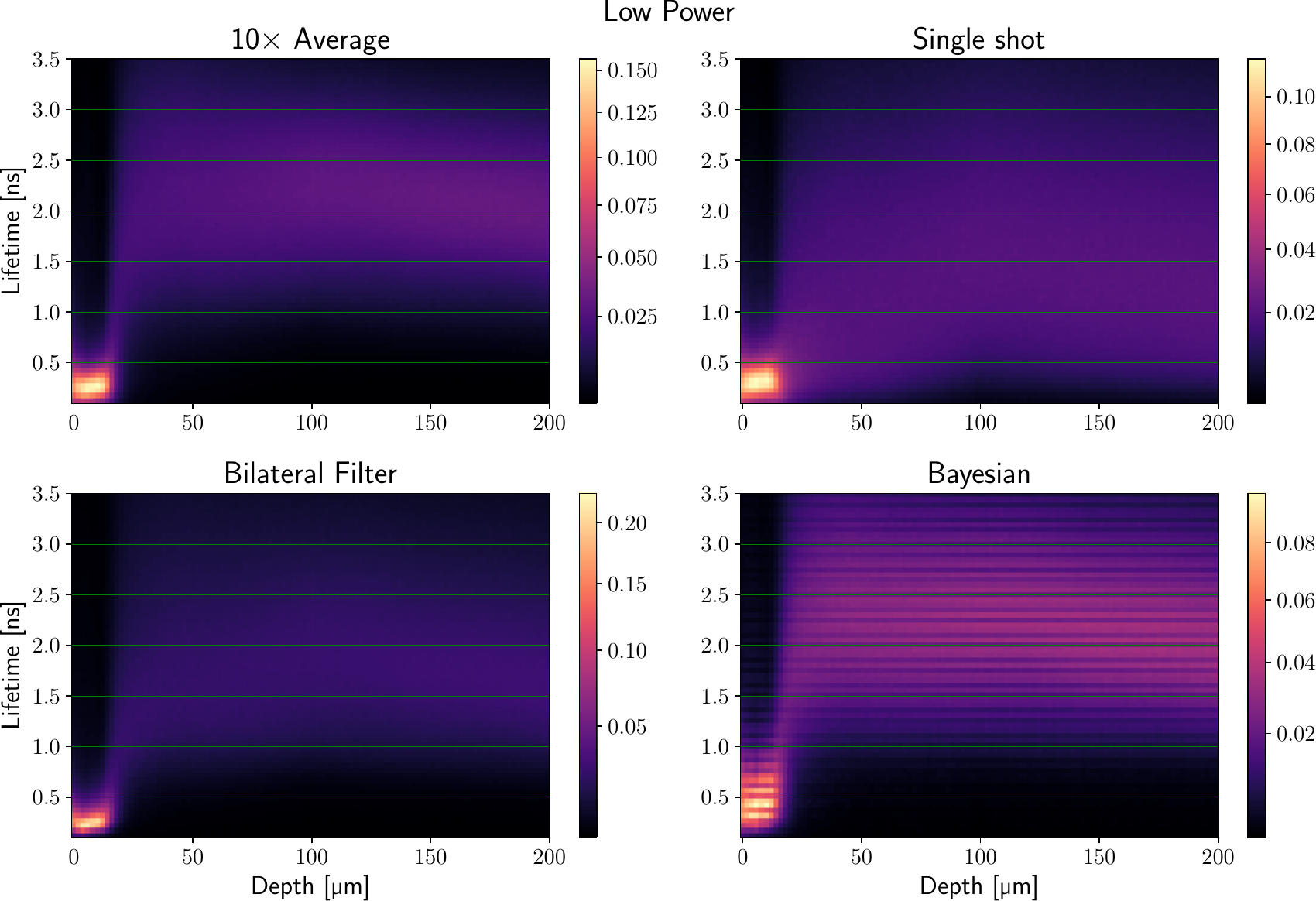}
\caption{
\textbf{Signal lifetime distribution with depth in kidney imaging.}
  Intensity-weighted histograms of image stacks corresponding to data shown in Fig.~\ref{fig:kidney}.
}
\label{fig:kidneyhistograms}
\end{figure}

Fig.~\ref{fig:kidneyhistograms} shows the lifetime histograms with depth at two locations, expanding on Fig.~\ref{fig:kidney} in the main text.
At one location marginally higher power was used, although this caused minor photodamage at shallower layers.
This was observed occasionally on the brighter (green) tubules and their puncta exhibiting $\sim$1\,ns lifetime, resulting in patches of longer lifetime $\sim$1.7\,ns (light blue).
This was associated with dilation or deformation of the tubules on the subsequent averaged scans due to higher exposure (lower power was used in averaged scans).

\clearpage

\section{Pulse Sampling Lifetime in Multiple Spectral Channels}\label{supp:multichannel}

\begin{figure}[!htbp]
\centering
\includegraphics[width=0.95\normaltextwidth]{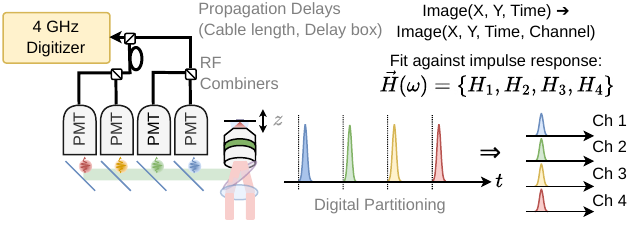}
\caption{
\textbf{Experimental setup showing multichannel configuration.}
}
\label{fig:setup}
\end{figure}

\begin{figure}[!htbp]
\centering
\includegraphics[width=1\normaltextwidth]{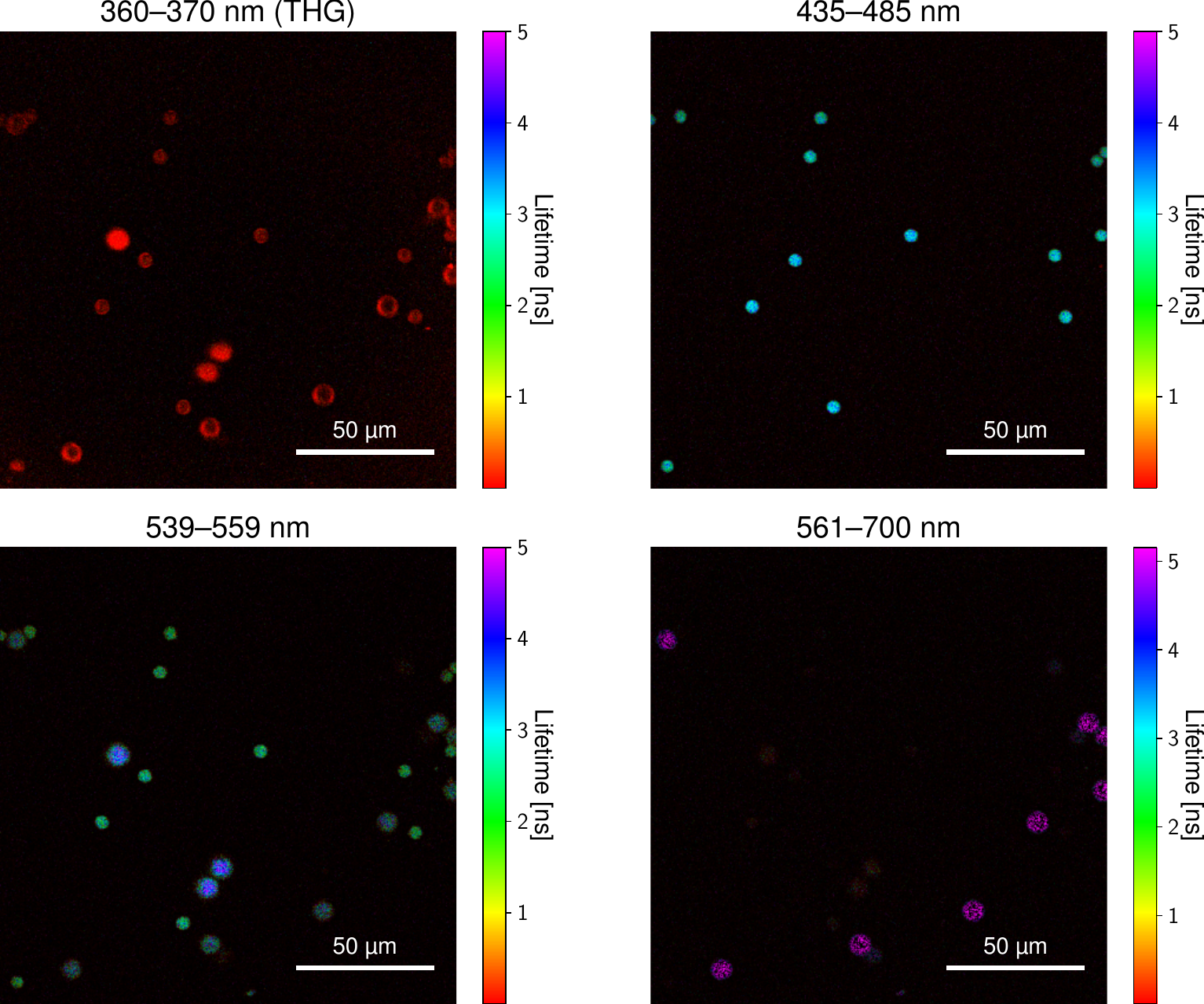}
\caption{
\textbf{Multispectral lifetime imaging.}
}
\label{fig:multichannel}
\end{figure}

Here we demonstrate the ability to do single-shot FLIM on a multichannel system.
As we discuss in the main text, extending fluorescence imaging from intensity-only to lifetime only requires fast digitization and hardware synchronization.
Many microscopes, however, use multiple channels, most commonly to spectrally separate fluorophores by emission wavelength.
This is done by digitizing multiple detector outputs.
Instead of using multiple digitizers, as a cost-saving measure, here we perform standard time multiplexing to incorporate multiple detectors and hence multiple channels.
In our setup, four PMTs were used as described in the \hyperref[methods:system]{Methods}.
Combining came with the cost of reduced transmission (approximately 33\%), which could in principle be mitigated by modestly amplifying each individual PMT, if this was found to be advantageous in terms of signal to noise.

The setup was built as a 4-channel microscope, although the focus of this work is on single-channel imaging.
The 4 channels were time-multiplexed by using three combiners (Mini-Circuits, ZSC-2-1W+) in a triangular configuration.
As a result, all four PMT outputs were directed to one digitizer channel and read out in serial.
Single traces of longer duration were thus partitioned into four channels by digitally slicing the output.
Delays were using pairs of different length coaxial cables (4 and 14\,ft) in addition to a delay box (set to 24\,ns).
Each trace segment corresponded to the signal contributed by one PMT.

The sample being imaged is a mixture of three fluorescent beads of various sizes (4.95, 7.32 and 9.77\,{\textmu}m) and with different emission central wavelengths (450, 600 and 520\,nm respectively) (Bangs Labs UMGB001, UMDG003, FCSY007).
The outputs of four PMTs, corresponding to the spectral filters seen in the figure, were combined using three RF power combiners before digitization, and read out in one digitizer channel to make use of the full sampling rate.
After digitization, the combined trace was then split into four channels and subsequently fit for lifetime.
The response of each PMT was calibrated.
Besides the splitters and digital slicing and padding or the digitizer readouts, the methods to obtain the lifetime image were the same as those described in the main text.
The two splitters in serial accounted for approximately 66\,\% transmission loss of the signal from the PMT.
This loss could in principle be compensated by having an amplifier after each PMT.

\end{appendices}